\documentclass[prb,twocolumn,tbtags,superscriptaddress,floatfix]{revtex4-2}

\usepackage{amsmath}
\usepackage{amssymb}
\usepackage{graphics}
\usepackage{epsfig}
\usepackage{verbatim}
\usepackage{textcomp}
\usepackage[colorlinks=true,urlcolor=blue,linkcolor=blue,citecolor=blue]{hyperref}
\usepackage{tabularx}
\usepackage{xcolor}
\usepackage{braket} 
\usepackage{float}
\usepackage{comment}

\usepackage{amsmath}
\usepackage{amssymb}
\usepackage{amsthm}

\begin{document}
	
\title{Alternative fast quantum logic gates using nonadiabatic Landau-Zener-St\"{u}ckelberg-Majorana transitions}

\author{A.~I.~Ryzhov}
\email[e-mail: ]{ryzhov@ilt.kharkov.ua}
\affiliation{B.~Verkin Institute for Low Temperature Physics and Engineering, Kharkiv
	61103, Ukraine}
\affiliation{Theoretical Quantum Physics Laboratory, Cluster for Pioneering Research, RIKEN, Wakoshi, Saitama, 351-0198, Japan}
\author{O.~V.~Ivakhnenko}
\affiliation{B.~Verkin Institute for Low Temperature Physics and Engineering, Kharkiv
	61103, Ukraine}
\affiliation{Theoretical Quantum Physics Laboratory, Cluster for Pioneering Research, RIKEN, Wakoshi, Saitama, 351-0198, Japan}
\author{S.~N.~Shevchenko}
\affiliation{B.~Verkin Institute for Low Temperature Physics and Engineering, Kharkiv
	61103, Ukraine}
\author{M.~F.~Gonzalez-Zalba}
\affiliation{Quantum Motion, 9 Sterling Way, London N7 9HJ, United Kingdom}
\author{Franco~Nori}
\affiliation{Theoretical Quantum Physics Laboratory, Cluster for Pioneering Research, RIKEN, Wakoshi, Saitama, 351-0198, Japan}
\affiliation{Quantum Computing Center, RIKEN, Wakoshi, Saitama, 351-0198, Japan}
\affiliation{Physics Department, The University of Michigan, Ann Arbor, MI 48109-1040, USA}

\begin{abstract}
	
	A conventional realization of quantum logic gates and control is based on resonant Rabi oscillations of the occupation probability of the system.
	This approach has certain limitations and complications, like counter-rotating terms. We study an alternative paradigm for implementing quantum logic gates based on Landau-Zener-St\"{u}ckelberg-Majorana (LZSM) interferometry with non-resonant driving and the alternation of adiabatic evolution and non-adiabatic transitions. Compared to Rabi oscillations, the main differences are a non-resonant driving frequency and a small number of periods in the external driving. We explore the dynamics of a multilevel quantum system under LZSM drives and optimize the parameters for increasing single- and two-qubit gates speed. We define the parameters of the external driving required for implementing some specific gates using the adiabatic-impulse model. The LZSM approach can be applied to a large variety of multi-level quantum systems and external driving, providing a method for implementing quantum logic gates on them.
	
\end{abstract}

\date{\today }
\maketitle


\section{Introduction}
\label{Sec:Introduction}

The conventional way of qubit state control is realized with resonant driving, resulting in Rabi oscillations (see, e.g., \cite{Krantz2019,Gu2017,Kockum2019,Jones2021}). There, the frequency of operation, the Rabi frequency, is defined by the driving amplitude; and so increasing the speed of operations means increasing the driving amplitude. This presents several challenges \cite{Yang2017b, Campbell2020}, including leakage to levels that lie outside the qubit subspace, breakdown of the rotating-wave approximation, and increased environmental noise. Instead of discussing the technological complications of the Rabi approach, let us consider here an alternative approach, based on a different paradigm of driving quantum systems.

When a quantum system exhibits an avoided-level crossing and is strongly driven, it can be described by the model  originally developped in several publications in 1932 and known as Landau-Zener-St\"{u}kelbeg-Majorana (LZSM) transitions (see, e.g., \cite{Shevchenko2010,Ivakhnenko2023,He2023,Glasbrenner2023,Weitz2023} and references therein). Effectively, the model can be split into two evolution stages: non-adiabatic transitions between the energy levels in the vicinity of the anti-crossing and adiabatic evolution far from the anti-crossing.

The energy-level occupation probabilities, as well as the relative phase between them, can be chosen by varying the driving parameters, providing a different paradigm for qubit state control \cite{Ivakhnenko2023,Chen2022}.

The LZSM transitions provide an alternative to conventional  gates based on resonant Rabi oscillations \cite{Kwon2021}. The energy level avoided crossing of a single qubit or two coupled qubits allows to controllably change states of such systems \cite{Quintana2013,Li2019,Zhang2013b} and to realize single- and two-qubit logic operations \cite{Campbell2020}. Recently, it was studied theoretically \cite{Benza2003, Hicke2006, Wei2008, Caceres2023} and demonstrated experimentally \cite{Cao2013, Wang2016a, Campbell2020,
	Zhang2021,Abadir1993,Heide2019,Heide2020,Heide2021a} that the LZSM model has several advantages over conventional gates based on Rabi oscillations.
These advantages include ultrafast speed of operation \cite{Cao2013, Cong2015}, robustness \cite{Hicke2006}, using baseband pulses (alleviating the need for pulsed-control signals) \cite{Campbell2020}, and reducing the effect of environmental noise \cite{Zhang2021}.

In this work we further develop the paradigm of the LZSM quantum logic gates. We investigate the single- and two-qubit systems' dynamics under an external drive
numerically solving the Liouville-von Neumann equation using the QuTiP framework \cite{Johansson2012,Johansson2013}. We explore the ways of finding the parameters for any arbitrary quantum logic gate with LZSM transitions and optimize the speed and fidelity of the quantum logic gates. We demonstrate the implementation of single-qubit X, Y, Hadamard gates and two-qubit iSWAP and CNOT gates using the LZSM transitions.

This paper is organized as follows. In Sec.~\ref{Sec:Hamiltonian and bases} we describe the qubit Hamiltonian and two main bases. In Sec.~\ref{Sec:Single-qubit gates} we demonstrate
X, Y, Hadamard, and phase gates implementations using both Rabi oscillations and LZSM transitions. We compare the speed and fidelities achieved with both paradigms.
We explore the way of increasing the gate speed and fidelity of the LZSM gates by using multiple transitions.
In Sec.~\ref{Sec:Two-qubit gates} we generalize the considered paradigm of using the adiabatic-impulse model for realization of quantum logic gates for multi-level quantum systems, and describe the realization of a two-qubit iSWAP gate with two LZSM transitions. The details for implementing other two-qubit gates, in particular a CNOT gate, are provided in Appendices~\ref{Sec:AppendixA} and ~\ref{Sec:AppendixB}. Sec.~\ref{Sec:Conclusion} presents the conclusions.

\section{Hamiltonian and bases}
\label{Sec:Hamiltonian and bases}

\begin{figure}[t]
	\includegraphics[width=0.8 \columnwidth]{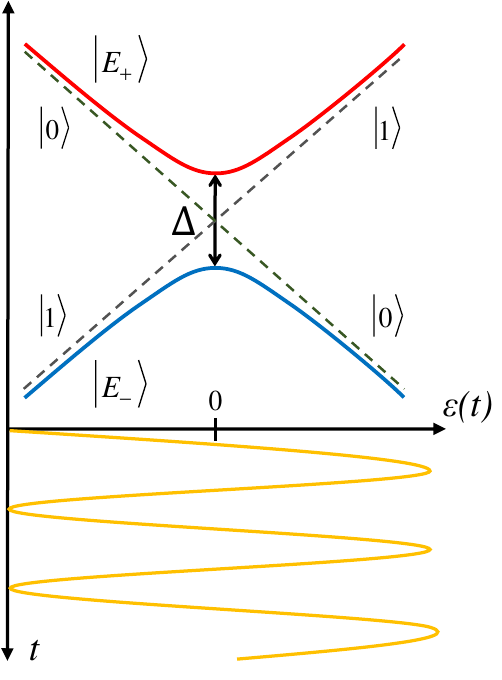}
	\caption{Energy diagram of a quantum two-level system under a periodic drive
		$\protect\varepsilon(t)$. Energy levels structure with two
		crossed diabatic levels $\ket{0}$, $\ket{1}$ and two adiabatic levels $\ket{E_\pm}$ with the an avoided crossing as a function of an energy detuning $\varepsilon(t)$.}
	\label{Fig:2LevelSystem}
\end{figure}

Consider the typical Hamiltonian for a driven quantum two-level system 
\begin{equation}
	\mathcal{H}(t)=\frac{\Delta }{2}\sigma _{x}+\frac{\varepsilon (t)}{2}\sigma
	_{z}= \frac{1}{2}\left( 
	\begin{array}{cc}
		\varepsilon(t) & \Delta \\ 
		\Delta & -\varepsilon(t)%
	\end{array}
	\right) ,  \label{H(t)}
\end{equation}%
where $\varepsilon(t)$ is the driving signal and $\Delta$ is the minimal energy gap between the two levels.
Here we consider the harmonic driving signal
\begin{eqnarray}
	\varepsilon (t)=A\sin \omega t. \label{Driwing_signal}
\end{eqnarray}%
The wave function is a superposition of two states of a quantum two-level system: 
\begin{equation}
	\left\vert\psi\right\rangle=\alpha(t)\left\vert0\right\rangle+\beta(t)\left
	\vert1\right\rangle=
	\begin{pmatrix}
		\alpha(t) \\ 
		\beta(t)
	\end{pmatrix}.
\end{equation}
The two main bases are: the \textit{diabatic} one, with diabatic energy levels $\{\ket{0}$, $\ket{1}\}$, where
the Hamiltonian becomes diagonalized when $\Delta=0$, and the \textit{adiabatic} basis $\ket{E_\pm}$, representing the eigenvalues of the total Hamiltonian, see Fig.~\ref{Fig:2LevelSystem}.
The relation between these bases is given by
\begin{subequations} 
	\begin{eqnarray}
		\left\vert E_{\pm }(t)\right\rangle &=&\gamma _{\mp }\left\vert
		0\right\rangle \mp \gamma _{\pm }\left\vert 1\right\rangle,
		\label{eigenstates}
	\end{eqnarray}
	where
	\begin{eqnarray}
		\gamma _{\pm } &=&\frac{1}{\sqrt{2}}\sqrt{1\pm \frac{\varepsilon (t)}{\Delta
				E(t)}}.  \label{gammas}
	\end{eqnarray}
\end{subequations}

Hereinafter, all the matrices of quantum logic gates, rotations $R_{x,y,z}$, matrices of adiabatic evolution $U$, and diabatic transition $N$ should be assumed to be represented in the adiabatic basis, while the Hamiltonians will be represented in the diabatic one. 

The dynamics of the quantum system with relaxation and dephasing can be described by the Lindblad equation. For simplicity, we consider the dynamics without relaxation and dephasing, described by the Liouville-von Neumann equation
\begin{equation}
	\frac{d\rho}{dt} = -\frac{i}{\hbar}\bigl[H(t),\rho\bigr],
	\label{VonNeumann}
\end{equation}
which coincides with the Bloch equations in the case of a two-level system.

\section{Single-qubit gates}
\label{Sec:Single-qubit gates}

We will describe a basic set of single-qubit gates (Sec.~\ref{Basic set of single-qubit gates}) and then explain how these can be performed using both the Rabi approach (Sec.~\ref{Sec:Rabi-based single-qubit operations}) and the LZSM approach (Sec.~\ref{Sec:Single-qubit operations based on LZSM transitions}). 

\subsection{Basic set of single-qubit gates}
\label{Basic set of single-qubit gates}
We consider different gates \cite{Kaye2006}: $X, Y, Z$ gates, phase gate $R_z(\phi)$, and the Hadamard gate $H$; which we write down here: 
\begin{flalign}
	&X=\sigma _{x}=R_{x}(\pi )=%
	\begin{pmatrix}
		0 & 1 \\ 
		1 & 0%
	\end{pmatrix}%
	, \\
	&Y=\sigma _{y}=R_{y}(\pi )=%
	\begin{pmatrix}
		0 & -i \\ 
		i & 0%
	\end{pmatrix} \Leftrightarrow
	R_{z}(\pi )R_{x}(\pi )
	, \\
	&Z=\sigma _{z}=R_{z}(\pi )=%
	\begin{pmatrix}
		1 & 0 \\ 
		0 & -1%
	\end{pmatrix}%
	, \\
	&  P(\phi) \equiv R_z(\phi)=%
	\begin{pmatrix}
		1 & 0 \\ 
		0 & e^{i\phi }%
	\end{pmatrix}%
	\Leftrightarrow
	\begin{pmatrix}
		e^{-i\phi /2} & 0 \\ 
		0 & e^{i\phi /2}%
	\end{pmatrix} \label{Phase_Gate}
	, \\
	&H=R_{y}(\pi /2)R_{z}(\pi)=\sqrt Y Z=\frac{1}{\sqrt{2}}%
	\begin{pmatrix}
		1 & 1 \\ 
		1 & -1%
	\end{pmatrix} \label{Hadamard_gate}
	,
\end{flalign}
where $R_{x,y,z}$ describes the rotations around the respective axes:
\begin{equation}
	\begin{gathered}
		R_{x,y,z}(\phi ) = \exp{\left(-i\sigma _{x,y,z}\frac{\phi }{2}\right)}= \\
		= \cos \left( \frac{\phi }{2} 	\right) I+i\sin \left( \frac{\phi }{2}\right) X,Y,Z.
	\end{gathered}
\end{equation}
Since the global phase of the density matrix $\rho$ is irrelevant and the dynamics is invariant to the multiplication of the density matrix $\rho$ by any complex number from the unit circle $e^{i\varphi}$,
the gate operator $e^{i\varphi} G$ is equivalent to the gate operator $G$, which we denote as
\begin{equation}
	e^{i\varphi}G\Leftrightarrow G.
	\label{gate_equivalence}
\end{equation}

\subsubsection{Phase gate $R_z(\phi)$}

The first gate we consider is the phase gate $R_z(\phi)$ in Eq.~\eqref{Phase_Gate}, which corresponds to a rotation around the $z$-axis by an angle $\phi$. To perform this gate, there is no need of a drive. Without driving, $\varepsilon=\text{const}$, the Bloch vector experiences a free natural rotation around the $z$-axis (which is the analogue to spin rotating in a magnetic field: the Larmor precession with frequency $\Omega_\text{L}$). The frequency of this free rotation depends on the distance between the energy levels
\begin{equation}
	\hbar \Omega_\text{L}=\Delta E=\sqrt{\varepsilon(t)^2+\Delta^2}.
	\label{LarmorFrequency}
\end{equation}
For a rotation by an angle $\phi$, a qubit needs to precess for a time
\begin{equation}
	t_{R_z}(\phi)=\frac{\phi}{\Omega_\text{L}}.
	\label{t_phase_gate}
\end{equation}
Since in the Rabi-based approach the energy detuning of the qubit
during the phase gate is
at the level anti-crossing $\varepsilon=0$, and in the LZSM-based approach it can be far away from the anti-crossing, the time of the phase gate in the LZSM-based approach can be reduced.
This difference in duration of the phase gates is demonstrated in Fig.~\ref{Fig:RabiGates}(a) and Fig.~\ref{Fig:LZSMGates}(a).

\subsection{Rabi-based single-qubit operations}
\label{Sec:Rabi-based single-qubit operations}

To perform any quantum logic gate with changing level occupation probability, the qubit should be excited by a time-dependent energy detuning $\varepsilon(t)$. A conventional way to achieve this is via Rabi oscillations with small amplitude $A\ll\Delta$ and with the qubit resonant frequency ($\hbar \omega = \Delta$), which we will compare with LZSM transitions with large amplitude $A>\Delta$ and non-resonant driving frequency $\omega$.

\begin{figure}[t]
	\center{\includegraphics[width=0.957							\columnwidth]{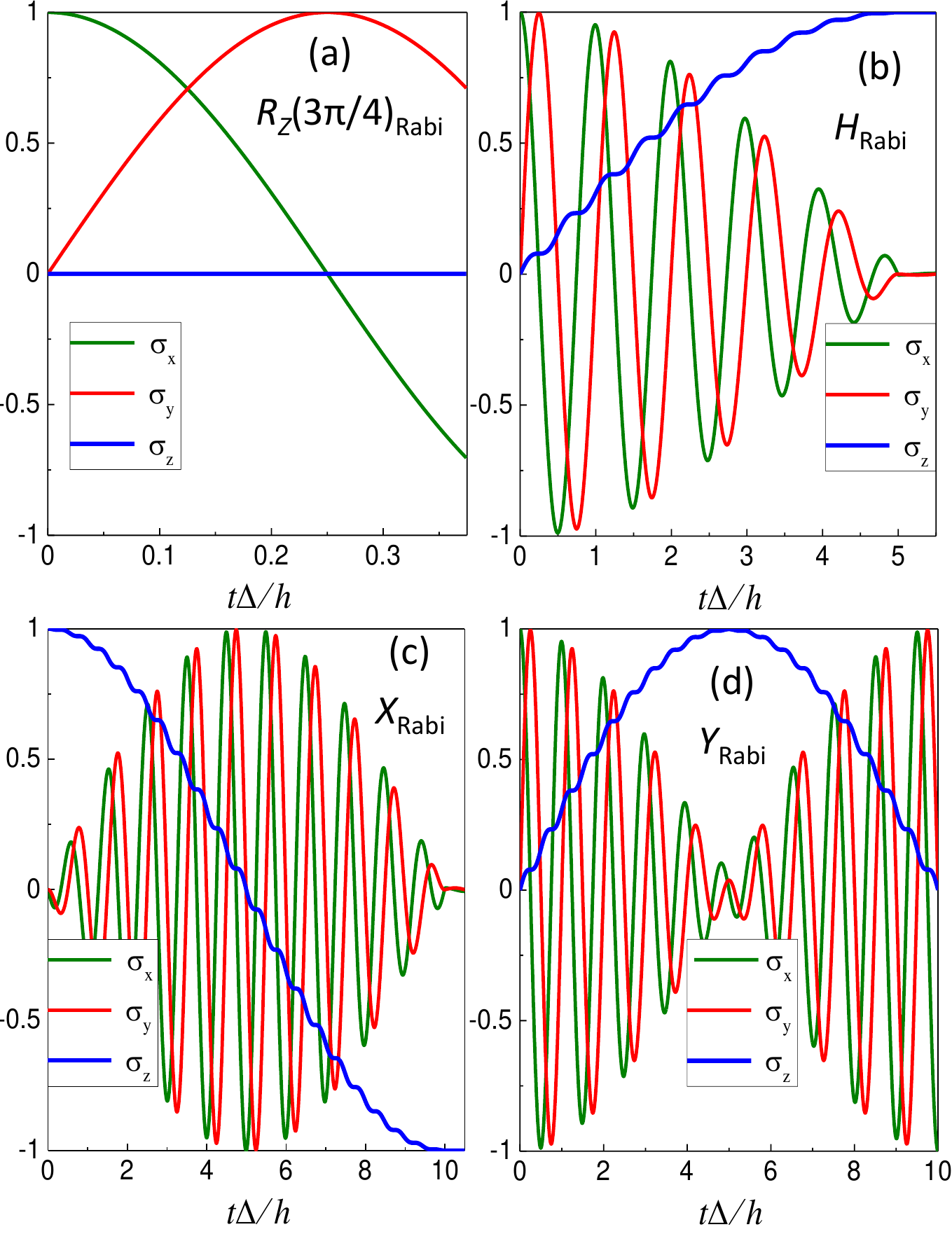}}
	\caption{Rabi-based single-qubit operations. Dynamics of the Bloch vector components in the adiabatic basis, obtained by solving the Liouville-von Neumann equation \eqref{VonNeumann}.
		(a) Phase operation $R_z({3\protect\pi}/{4})$ with a superposition initial state $\protect\psi(t_\text{i})=(\ket{0}+\ket{1})/\sqrt{2}\protect$.
		(b) Hadamard operation $H$ with superposition initial state.
		(c) $X$ operation with the ground-state initial conditions.
		(d) $Y$ operation with the superposition initial state.
		For (b,c,d) the driving frequency is resonant $\protect\hbar\omega=\Delta$, the
		amplitude is small $A=0.1\Delta$. For (b) the number of periods of the resonant drive is $N_\text{e}=5$, for (c,d) $N_\text{e}=10$.
		The amplitude is defined by the number of periods of the external drive $A=\Delta / N_\text{e}$.
	}
	\label{Fig:RabiGates}
\end{figure}

\begin{figure}[t]
	\center{\includegraphics[width=1							\columnwidth]{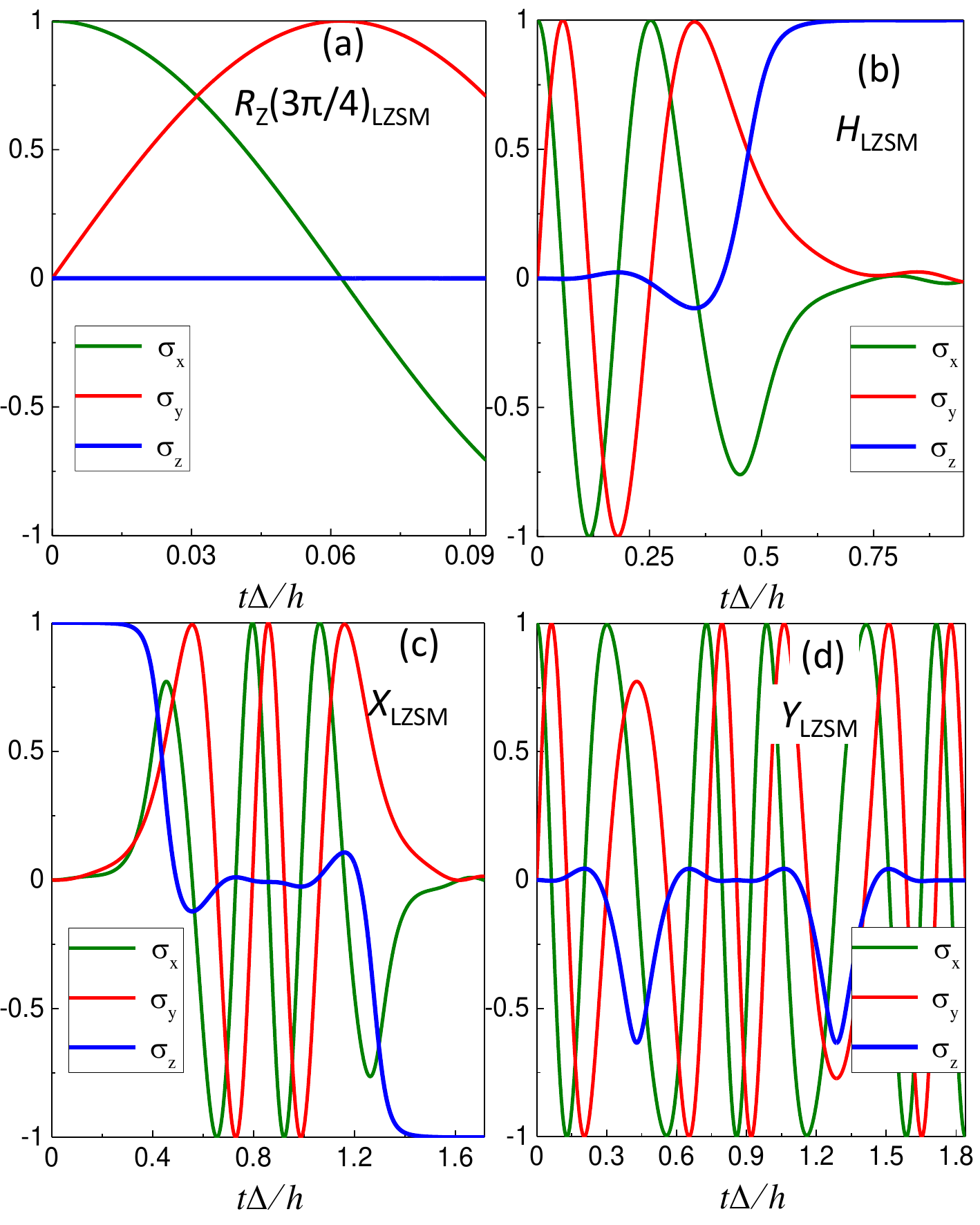}}
	\caption{LZSM-based single-qubit operations. Dynamics of the Bloch vector components in the adiabatic basis, obtained by solving the Liouville-von Neumann equation \eqref{VonNeumann}.
		(a) Phase operation $R_z({3\protect\pi}/{4})$ with the superposition initial state $\protect\psi(t_\text{i})=(\ket{0}+\ket{1})/\sqrt{2}\protect$.
		(b) Hadamard operation $H$ with superposition initial state.
		(c) $X$	operation with ground-state initial conditions.
		(d) $Y$ operation with the superposition initial state.
		For (b,c,d) the parameters are chosen so that $\mathcal{P}=0.5$. For (b) the amplitude $A=4.31\Delta$, for (c) and (d) $A=3.8872\Delta$. Compared with the respective gates based on Rabi oscillations,
		Fig.~\ref{Fig:RabiGates}, LZSM-based gates are much faster.}
	\label{Fig:LZSMGates}
\end{figure}

Here we describe how the single-qubit operations are implemented with Rabi
oscillations and demonstrate the dynamics of the Bloch sphere coordinates for several logic gates in Fig.~\ref{Fig:RabiGates}.
Rabi oscillations occur during the resonant driving at $\delta\omega=\omega-\omega_\text{q} \ll\omega$ (where $ \omega_\text{q} = \Delta E / \hbar \approx \Delta / \hbar$  is the qubit resonant frequency) with small amplitude $A\ll\Delta$, and harmonic driving
signal, Eq.~\eqref{Driwing_signal}.

The Rabi oscillations lead to a periodic change of
the level occupation with Rabi frequency
\begin{eqnarray}
	\Omega_\text{R}=\frac{A\Delta}{2\hbar\Delta E}\approx\frac{A}{2\hbar}.
	\label{Rabi_frequency}
\end{eqnarray}
During the oscillations, the $z$-component of the Bloch vector changes as $z(t)=\cos\Omega_\text{R}t$, when the initial state is the ground state $\ket{E_-}$. While the state probability is evolving, a phase change also occurs with frequency
\begin{equation}
	\hbar \Omega_L \approx \Delta.
\end{equation}
We define the Rabi oscillations evolution as a combination of two
rotations
\begin{equation}
	U_\text{Rabi}(t)=
	R_z\left(\Omega_\text{L}t\right)R_x\left(\Omega_\text{R}t\right).
	\label{Rabi_Dynamics}
\end{equation}
Using Eq.~\eqref{Rabi_frequency}, we can rewrite it as
\begin{equation}
	U_\text{Rabi}(t)=R_z\left(\Omega_\text{L}t\right) R_x \biggl( \frac{A}{2 \hbar} t \biggr) = R_z\left(\Omega_\text{L}t\right) R_x \biggl( \frac{S}{2 \hbar}\biggr),
	\label{Rabi_Dynamics2}
\end{equation}
which shows that the angle of rotation around the $x$-axis is proportional to the area $S=At$ under the envelope of the Rabi pulse.

When there is no phase difference between rotations, $\Omega_\text{L} t  = \Omega_\text{R} t + 2 \pi n$, we obtain
\begin{equation}
	U_\text{Rabi}(t)=
	R_z\left(\Omega_\text{L}t\right)R_x\left(\Omega_\text{R}t\right)=R_y(\Omega_\text{R}t),
\end{equation}
and the Rabi evolution results in  a rotation around the $y$-axis.

To perform an $X$ operation, we drive the system by Rabi pulses during a time $T_\text{R}$, so that the area under the envelope is
\begin{equation}
	S = A T_\text{R} = 2 \pi .
	\label{Rabi_amplitude}
\end{equation}
In order for the driving to end at zero amplitude, we take an integer number of periods of the sine.

After that, we need to change the phase to obtain an $X$ operation from a $Y$ rotation, so we perform the $R_z$ rotation by idling the drive for a time $T_\text{I}$ 
with the condition
\begin{equation}
	\Omega_\text{L}(T_\text{I}+T_\text{R})=2\pi n,
\end{equation}
then finally the $X$ gate is realized as
\begin{equation}
	\begin{gathered}
		R_z(\Omega_\text{L}T_\text{I}) U_\text{Rabi}(T_\text{R})= \\
		= R_z(\Omega_\text{L}T_\text{I}) R_z\left(\Omega_\text{L}T_\text{R}\right)R_x\left(\Omega_\text{R}T_\text{R}\right) =R_x(\pi) = X,
	\end{gathered}
\end{equation}
see Fig.~\ref{Fig:RabiGates}(c).
To perform the Hadamard gate, we need to apply the Rabi pulse
with the duration twice shorter than for the $X$ gate,  $T_\text{R}={\pi}/{\Omega_\text{R}}$, with the condition on the idling time $T_\text{I}$:
\begin{equation}
	\Omega_\text{L}(T_\text{I}+T_\text{R})=\pi + 2\pi n.
\end{equation}
As a result, we obtain the Hadamard gate as
\begin{equation}
	\begin{gathered}
		R_z(\Omega_\text{L}T_\text{I}) U_\text{Rabi}(T_\text{R})= \\
		= R_z(\Omega_\text{L}T_\text{I}) R_z\left(\Omega_\text{L}T_\text{R}\right)R_x\left(\Omega_\text{R}T_\text{R}\right) = \\
		= R_z(\pi) R_x(\pi/2) =  H,
	\end{gathered}
\end{equation}
which is demonstrated in Fig.~\ref{Fig:RabiGates}(b).


\subsubsection{Gaussian envelope optimization for Rabi-based gates}

Since the Rabi-oscillations model assumes small amplitudes of the driving signal, to increase the gate fidelity, a small amplitude at the start and end points should be used. To achieve a high gate speed, a large driving amplitude $A$ between these points should be used. Hence, to increase the gate speed, now we use a Gaussian-shaped envelope $A(t)$ for the driving signal
\begin{eqnarray}
	\varepsilon (t)=A(t)\sin \omega t.
\end{eqnarray}
We now consider a Rabi pulse with duration $T_R$ with the envelope in the form 
\begin{equation}
	\begin{gathered}
		A(t)=
		\begin{cases}
			A_{0}\exp{\left[-\frac{(t-\tau)^2}{2\sigma_\text{G}^2}\right]}, \ \ \ t<T_\text{R}\\
			0, \ \ \ t>T_\text{R}
		\end{cases}
	\end{gathered}
\end{equation}
with the tails of the Gaussian distribution truncated at some distance $G$ from the peak, normalized to the standard deviation $\sigma_\text{G}$,
\begin{eqnarray}
	G=\frac{T_\text{R}}{2 \sigma_\text{G}}
	\label{G_Gaussian}
\end{eqnarray}
and the peak of the distribution at time
\begin{eqnarray}
	\tau=\frac{T_\text{R}}{2}.
\end{eqnarray}

The angle of rotation around the $x$-axis in Eq.~\eqref{Rabi_Dynamics2} is defined by the area under the envelope of the Rabi pulse. 
For the $X$ operation it is given by Eq.~\eqref{Rabi_amplitude}.
So the area under the truncated Gaussian distribution should be the same as for the original signal with constant amplitude and rectangular shape of the pulse.
This condition determines the amplitude of the distribution as
\begin{eqnarray}
	A_\text{0}=\sqrt{\frac{2 \pi}{\sigma_\text{G} S_\text{G}}},
\end{eqnarray}
where $S_\text{G}$ is the normalized area of the truncated Gaussian distribution
\begin{eqnarray}
	S_\text{G}= \frac{1}{\sqrt{2 \pi}} \int_{-G}^{G}e^{-\frac{x^2}{2}}dx.
\end{eqnarray}

\subsection{Single-qubit operations based on LZSM transitions}
\label{Sec:Single-qubit operations based on LZSM transitions}

Here we describe how to implement single-qubit operations based on LZSM
transitions using the adiabatic-impulse model (AIM), also known as the transfer-matrix method, and demonstrate the dynamics of the Bloch sphere coordinates for several logic gates in Fig.~\ref{Fig:LZSMGates}, that can be compared with the dynamics of the same gates realized with Rabi oscillations in Fig.~\ref{Fig:RabiGates}. For the diabatic LZSM transitions,
we need the following approximations: $A>\Delta$ and $2\pi/\omega < t_\text{trans}$, where $t_\text{trans}$ is the transition time. After that time the result of the adiabatic-impulse model will asymptotically coincide with the exact dynamics  ~\cite{Vitanov1999,Ivakhnenko2023}.

\subsubsection{Adiabatic-impulse model. Single-passage drive}

In the adiabatic-impulse model, the time evolution is considered as a combination of adiabatic
(non-transition) and diabatic (transition) evolutions.
The adiabatic evolution is described by
the adiabatic time-evolution matrix, 
\begin{equation}
	U(t_i,t_j)= 
	\begin{pmatrix}
		e^{-i\zeta (t_i,t_j)} & \ 0 \\ 
		0 & \ e^{i\zeta (t_i,t_j) }%
	\end{pmatrix}
	=e^{-i\zeta \sigma _{z}}=R_z(2\zeta),  \label{AdiabaticEvolution}
\end{equation}
where $\zeta (t_i,t_j)$ is the phase accumulated during the
adiabatic evolution 
\begin{equation}
	\zeta (t_i,t_j)=\frac{1}{2\hbar }\int_{t_i}^{t_j} \Delta E(t) \ dt = \frac{1}{2\hbar}\int_{t_i}^{t_j}\sqrt{\varepsilon(t)^2+%
		\Delta^2}dt,
	\label{zeta}
\end{equation}
and $\Delta E(t) = E_+(t)  - E_-(t)$.
Then, the \textit{diabatic} evolution (transition) is described by the matrix 
\begin{equation}
	\begin{gathered}
		N= 
		\begin{pmatrix}
			Re^{-i \phi_\text{S}} & -T \\ 
			T & Re^{i \phi_\text{S}}%
		\end{pmatrix}%
		= \\
		=R_{z}\left( \phi_{\mathrm{S}}\right) R_{x}\left( \theta
		\right) R_{z}\left( \phi_{\mathrm{S}}\right),
		\label{single_qubit_diab_matr}
	\end{gathered}
\end{equation}
where
\begin{equation}
	\begin{gathered}
		T=\sqrt{\mathcal{P}},	\\
		R=\sqrt{1-\mathcal{P}}
		\label{tran_refl_coef}
	\end{gathered}
\end{equation}
are the transition and reflection coefficients, \begin{equation}
	\mathcal{P}=\exp{(-2\pi\delta)}
	\label{P_lzsm}
\end{equation}
is the LZSM probability of excitation of the qubit with a single transition from the ground state $\ket{E_-}$, $\delta={\Delta^2}/{4v}$ is the adiabaticity parameter, $v=\varepsilon^\prime(0)$ is the speed of the anti-crossing passage and
\begin{eqnarray}
	\phi _{\text{S}} &=&\frac{\pi }{4}+\delta (\ln {\delta }-1)+\mathrm{Arg}
	[\Gamma (1-i\delta ) ] \label{StokesPhase}
\end{eqnarray}
is the Stokes phase \cite{Ivakhnenko2023}. The $\theta$ angle can be found from the equation 
\begin{equation}
	\sin ^{2}(\theta /2)=\mathcal{P}.
	\label{P_to_theta}
\end{equation}
The \textit{inverse} transition matrix can be written as
\begin{equation}
	\begin{gathered}
		N^\text{inv}=N^\top =
		\begin{pmatrix}
			Re^{-i \phi_\text{S}} & T \\ 
			-T & Re^{i \phi_\text{S}}%
		\end{pmatrix}\Leftrightarrow
		\\
		\Leftrightarrow
		\begin{pmatrix}
			Re^{-i (\phi_\text{S} - \pi)} & -T \\ 
			T & Re^{i (\phi_\text{S} - \pi)}
		\end{pmatrix}.
	\end{gathered}
\end{equation}
The single transition evolution matrix in the general case with \textit{adiabatic} evolution matrix before the transition $U_1$ and after the transition $U_2$ is given by
\begin{equation}
	\begin{gathered}
		U_\text{LZSM}= U_2 N_1 U_1 = 
		\begin{pmatrix} 
			U^\prime_{11}& 	U^\prime_{12}
			\\
			-U^{\prime*}_{12} &  U^{\prime*}_{11} \end{pmatrix},
	\end{gathered} \label{Single_Transition_General_Matrix}
\end{equation}
where
\begin{eqnarray}
	U^\prime_{11}&=&R_1 \exp{ [ -i (\phi_\text{S1} + \zeta_{1} + \zeta_{2}) ] }, \notag \\
	U^\prime_{12}&=&- T_1 \exp{ [ i (\zeta_{1} - \zeta_{2}) ] },\notag	\\
	\zeta_{1} &=& \zeta(0,t_{N1}),\notag	\\
	\zeta_{2} &=& \zeta(t_{N1},t_{\text{final}}).\notag
\end{eqnarray}
Here, $t_{N1}$ is the time of the level anti-crossing passage, $t_{\text{final}}$ is the end time of the drive.

Then, we consider the same adiabatic evolution before and after the transition $\zeta=2\zeta_1=2\zeta_2$.
In that case, a single LZSM transition gate can be represented \cite{Sillanpaeae2006,Sillanpaeae2007} as a combination of rotations
\begin{equation}
	\begin{gathered}
		U_\text{LZSM}(\mathcal{P},\phi_\text{total})=R_z(\phi_\text{total}	)R_x(\theta)R_z(\phi_\text{total}), \\
		U^\text{inv}_\text{LZSM }(\mathcal{P},\phi_\text{total})=U_\text{LZSM}(\mathcal{P},\phi_\text{total}-\pi),
	\end{gathered}
	\label{U_LZSM}
\end{equation}
where $\phi_\text{total}=\phi_{\mathrm{S}}+\zeta$, and
$U^\text{inv}_\text{LZSM}$ corresponds to the inverse transition.
Using this LZSM gate, we can define a basic set of gates.

For an $X$ gate, the two-level system needs to perform a transition with probability $\mathcal{P}=\exp{(-2\pi\delta)}=1$; which means that the adiabaticity parameter $\delta={\Delta^2}/{4v}\rightarrow 0$, requiring an infinite speed of the anti-crossing passage $v=\varepsilon^\prime(0)\rightarrow\infty$ or a zero energy splitting $\Delta$. Hence, it is difficult  to implement the $X$ operation with high fidelity using only a single passage.
Therefore, at least two transitions are needed for implementing the $X$ gate with sufficient fidelity.

For the LZSM transition, we need to start and  finish
the evolution far from the anti-crossing region. So we now consider the
harmonic driving signal $\varepsilon(t)=-A\cos(\omega t)
$. This signal is linear in the anti-crossing region
\begin{equation}
	\frac{d\varepsilon}{dt}  \biggm \vert_{\varepsilon \approx 0}  \approx A \omega =  \text{const}.
	\label{linear_signal}
\end{equation}
We obtain a relation between the amplitude $A$ and the frequency $\omega$ for certain LZSM probability $\mathcal{P}$
\begin{equation}
	\mathcal{P} = \exp{\biggl[-2\pi\frac{\Delta^2}{4A\omega\hbar}\biggr]}
	\rightarrow 
	\omega=\frac{-\pi\Delta^2}{2A \hbar \ln \mathcal{P}}.  \label{AwLZSMConnection}
\end{equation}
Then, we find an amplitude which satisfies some value
of $\phi_\text{total}$ and $\mathcal{P}$,
\begin{eqnarray}
	\phi_\text{total}&=&\frac{\pi }{4}+\delta (\ln {\delta}-1)+\mathrm{Arg}%
	[\Gamma (1-i\delta)]+ \\
	&+&\frac{1}{2\hbar}\int_{0}^{\frac{\pi}{\omega}}\sqrt{\varepsilon(t)^2+%
		\Delta^2}dt,  \notag \label{PhitotalFormula}  \\
	\delta&=&-\frac{\ln \mathcal{P}}{2\pi},
	\label{delta}
\end{eqnarray}
where we used that the harmonic driving satisfies the initial conditions far from the anti-crossing region  $\varepsilon(t)=-A\cos{\omega t}$.

A single LZSM transition is convenient for implementing rotations to any angle $\theta<\pi$, for example $\theta={\pi}/{2}$, which is needed for the Hadamard gate. Following Eq.~\eqref{P_to_theta}, the angle $\theta={\pi}/{2}$ corresponds to the target probability of a single LZSM transition
\begin{equation}
	\mathcal{P}=\sin^2\left(\frac{\pi}{4}\right)=\frac{1}{2}.
\end{equation}
This LZSM transition is \textit{non-instantaneous}: the probability oscillates for some time after the transition, and the value of the upper-level occupation obtained from the formulae cannot be exactly reached until the end of the oscillations \cite{Vitanov1999}.
The parameters for the Hadamard gate implementation can be found from Eq.~\eqref{Single_Transition_General_Matrix} by equating $U_\text{LZSM}$ to the matrix of the gate~\eqref{Hadamard_gate}, and solving the system of equations
\begin{equation}
	\begin{aligned}
		&\mathcal{P}=\frac{1}{2} \rightarrow T=R=\frac{1}{\sqrt{2}},\\
		&\zeta_{1}-\zeta_{2} = \frac{\pi}{2}+\pi n_1,		\\ &\phi_\text{S}+\zeta_{1}+\zeta_{2}=\frac{\pi}{2}+\pi n_2.
	\end{aligned}
\end{equation}
From this system we obtain the total phase
\begin{equation}
	\phi_\text{total}=\pi n,
	\label{LZSMHadamardPhase}
\end{equation}
so the Hadamard gate can be presented as
\begin{equation}
	\begin{gathered}
		H=R_y(\pi/2) R_z(\pi)=U_\text{LZSM}\left(1/2, 2\pi n\right) R_z(\pi) = \\
		=R_z(\pi) U_\text{LZSM}\left(1/2, \pi+2\pi n\right),
		\label{HadamardtotphaseCond}
	\end{gathered}
\end{equation}
where $n$ is an integer. The dynamics of the Hadamard gate is shown in Fig.~\ref{Fig:LZSMGates}(b).

How to find the driving amplitude $A$ and frequency $\omega$ required for certain $\mathcal{P}$ and $\phi_\text{total}$ is described in Section~\ref{Optimized TM multiple Passage}.
After the transition is completed, to perform some rotation around the $z$-axis (phase gate), we need to apply a constant signal with the same energy detuning $\varepsilon$ as we had after completing the previous operation.

Alternatively, LZSM gates can also be realized with the position of the energy detuning before and after the gate at the level anti-crossing $\varepsilon=0$ \cite{Campbell2020}.

\subsubsection{Double-passage drive}

\begin{figure}[t]
	\center{\includegraphics[width=1\linewidth]{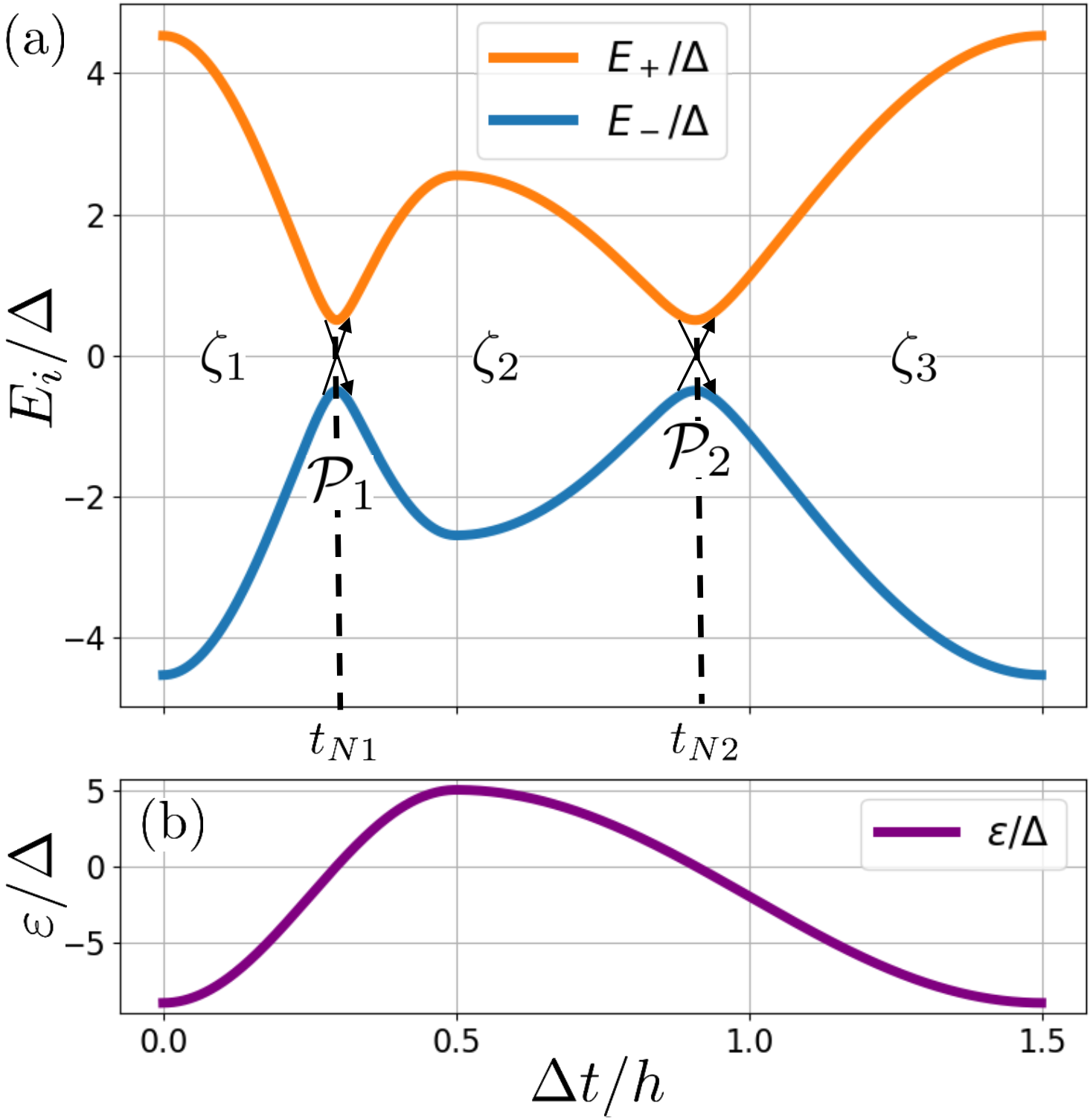}}
	\caption {Energy levels dynamics for a drive with two passages of the level anti-crossing.
		(a) Adiabatic energy levels versus time.
		(b) The energy detuning $\varepsilon$ versus time.
		Here, $\mathcal{P}_i$ is the value of the LZSM probability of the diabatic transition $N_i$ during the passage of the level anti-crossing at time $t_{N(i)}$.
		Here, $\zeta_i$ is the phase accumulation gained during the corresponding adiabatic evolution interval. }
	\label{Fig:Enegry-level_dynamics}
\end{figure}

Consider now an arbitrary external drive $\varepsilon(t)$ with two passages through the energy-level anti-crossing, linear in the anti-crossing region. The adiabatic energy levels as a function of time are illustrated in Fig.~\ref{Fig:Enegry-level_dynamics}(a).
We obtain the double transition evolution matrix in the general case:

\begin{equation}
	\Xi= U_3 N_2 U_2 N_1^\text{inv} U_1 = U^\text{inv}_\text{LZSM(1)} U_\text{LZSM(2)}=
	\begin{pmatrix} 
		\Xi_{11}& \Xi_{12}\\
		-\Xi_{12}^* & 
		\Xi_{11}^* \end{pmatrix}
\end{equation}\label{TwoPassagesMatrix}
where
\begin{eqnarray}
	\Xi_{11}&=&(R_1 R_2 e^{-i (\phi_\text{S1} + \phi_\text{S2} + 2\zeta_{2})} + T_1 T_2 ) e^{i (\zeta_2 - \zeta_1 - \zeta_3)}, \notag \\
	\Xi_{12}&=&(R_1 T_2  - T_1 R_2 e^{-i (\phi_\text{S1}+\phi_\text{S2} + 2\zeta_{2})}) e^{i (\phi_\text{S1}+\zeta_1+\zeta_2 - \zeta_3)},\notag	\\
	\zeta_{1} &=& \zeta(0,t_{N1}),\notag	\\
	\zeta_{2} &=& \zeta(t_{N1},t_{N2}),\notag	\\
	\zeta_{3} &=& \zeta(t_{N2},t_{\text{final}}) \notag
\end{eqnarray}

Here, $t_{N1}$ and $t_{N2}$ are the times of the first and second passages of the level anti-crossing respectively, $t_{\text{final}}$ is the end time of the drive, see Fig.~\ref{Fig:Enegry-level_dynamics}. By equating this evolution matrix to the matrix of the required quantum gate, one can find the parameters of the driving signal that implements this gate. For example, for an $X$ gate the driving signal should satisfy the conditions
\begin{equation}
	\begin{aligned}
		&\mathcal{P}_1 + \mathcal{P}_2 = 1 \rightarrow T_1 = R_2,	\\
		&\phi_\text{S1} + \phi_\text{S2} + 2 \zeta_2 = \pi + 2 \pi n_1,	\\
		&\phi_\text{S2} - \zeta_1 + \zeta_2 + \zeta_3 = \pi/2 + 2 \pi n_2.
	\end{aligned}
\end{equation}
To simplify the result, we consider a periodic driving with the same slope in the anti-crossing region during each transition $\varepsilon(0)\approx vt$, which means $\mathcal{P}_1=\mathcal{P}_2=1/2$, $T_1=T_2=R_1=R_2=1/\sqrt{2}$, and with the same adiabatic evolution between transitions $\zeta=\zeta_1=\zeta_2/2=\zeta_3$. 
After this simplification, we obtain a matrix of the double transitions with only two parameters \cite{Ivakhnenko2023}: adiabatic phase gain $\zeta$ and excitation probability $\mathcal{P}$, with
\begin{eqnarray}
	\Xi \equiv \sqrt{U_{2}}N^\text{inv}U_{1}N\sqrt{U_{2}}=
	\begin{pmatrix} 
		\Xi _{11} & \Xi _{12} \\ 
		-\Xi^* _{12} & \Xi _{11}^{\ast } \end{pmatrix},
	\label{SingleTransitionMatrix}
\end{eqnarray}%
where
\begin{subequations}\begin{eqnarray}
		\Xi _{11} &=&-R^2e^{-2i\Phi _{\mathrm{St}}}-T^2,
		\\
		\Xi _{12} &=&-\Xi _{12}^* =
		-2iRT\sin(\Phi_\mathrm{St}), \label{Xi_12St}\\
		\Phi _{\mathrm{St}}&\equiv&\phi_{\mathrm{S}}+\zeta \label{PhiSt},
\end{eqnarray}\end{subequations}
and $\Phi _{\mathrm{St}}$ is a St\"{u}ckelberg phase.
For the symmetric drive with $\mathcal{P}_1=\mathcal{P}_2$ and $\zeta=\zeta_1=\zeta_2/2=\zeta_3$, $\Phi _{\mathrm{St}}=\phi_\text{total}$.
For the $X$ operation, shown in Fig.~\ref{Fig:LZSMGates}(c), 
we used two LZSM transitions with LZSM probability $\mathcal{P}=1/2$, and total phase for each transition

\begin{equation}
	\phi_\text{total}=\frac{\pi}{2}+\pi n,
	\label{2pinPhaseCondition}
\end{equation} 
which is the condition for a constructive interference (see, e.g., \cite{Ivakhnenko2023}).
Indeed, using Eq.~\eqref{U_LZSM},
\begin{equation}
	U^\text{inv}_{\text{LZSM}}(\pi/2+\pi n, 1/2)U_\text{LZSM}(\pi/2+\pi n, 1/2)=R_x(\pi)=X.
\end{equation} 
In principle, a double LZSM transition drive, in conjunction with a rotation around the $z$-axis, can implement any single-qubit gate.

For the Hadamard gate implemented by two LZSM transitions with the same slope in the anti-crossing region during each transition ($\mathcal{P}_1=\mathcal{P}_2$) the driving signal should satisfy the conditions
\begin{eqnarray}
	\begin{aligned}
		&\mathcal{P}=\frac{2 \pm \sqrt{2}}{4}, \\		
		&\phi_\text{S}+\zeta_{2}=\frac{\pi}{2}+2\pi n_1,\\
		&\zeta_{1}+\zeta_{2}+\zeta_{3}=\frac{\pi}{2}+2\pi n_2,\\
		&\zeta_{1}-\zeta_{3}=2\pi n_3.
	\end{aligned}
	\label{H_two_passages_condition}
\end{eqnarray}

\subsubsection{Optimization to speed up gates: multiple passage drive }
\label{Optimized TM multiple Passage}
To speed up the gate, multiple LZSM transitions can be used. 
We consider the simplest drive
\begin{equation}
	\begin{gathered}
		\varepsilon(t) =
		\begin{cases}
			- A , \ \ \ \ \ \ \ \ \ \ 0 < t < t_\text{pre},	\\
			- A \cos \omega t  , \ \   t_\text{pre} < t < t_\text{pre} + \frac{2 \pi k }{\omega},	\\
			- A,  \ \ \ \ \ \ \ \ \ \  t_\text{pre} + \frac{2 \pi k }{\omega} < t < t_\text{pre} + \frac{2 \pi k }{\omega} + t_\text{after},
		\end{cases}
	\end{gathered}
	\label{LZSMdrive}
\end{equation}
with an even number $2k$ of successive LZSM transitions with the same probability of LZSM transition $\mathcal{P}$, and St\"{u}ckelberg phase $\Phi_\mathrm{St}$, and the idling periods with phase accumulation at the start and end of the drive with durations $t_\text{pre}$ and $t_\text{after}$, respectively. Here, $k=1,2,...$ is the number of periods of the cosine.

For the case of four LZSM transitions, the evolution matrix of the harmonic part of the drive, $\varepsilon(t)= - A \cos \omega t$, can be found as the multiplication of two evolution matrices for double passage ~\eqref{SingleTransitionMatrix}:
\begin{subequations} 
	\begin{eqnarray}
		\Xi_\text{Q}&=& \Xi^2 =
		\begin{pmatrix} \Xi _{\text{Q}11} & \Xi _{\text{Q}12} \\ 
			-\Xi^*_{\text{Q}12} & \Xi_{\text{Q}11}^{\ast } \end{pmatrix},
		\label{4transitions}
	\end{eqnarray}
	where
	\begin{eqnarray}
		\Xi_{\text{Q}11}&=&R^4e^{-4i\Phi_\text{St}}+T^4+\\&+&2R^2T^2\left[e^{-2i\Phi_\text{St}}-2\sin^2(\Phi_\text{St})\right], \notag\\
		\Xi_{\text{Q}12}&=&4iRT\sin(\Phi_\text{St})\left[R^2\cos(2\Phi_\text{St})+T^2\right]. \label{sigma_q12}
	\end{eqnarray}
\end{subequations}
Since $T=\sqrt{\mathcal{P}}$ and $R=\sqrt{1-\mathcal{P}}$, the evolution matrix depends on two parameters of the drive: the probability of a single LZSM transitions $\mathcal{P}$, and the St\"{u}ckelberg phase $\Phi_\mathrm{St}$. The idling periods before and after the main part of the drive result in the phase-shift gates $R_z(\phi_\text{pre})$ and $ R_z(\phi_\text{after})$, respectively [see Eq.~\eqref{Phase_Gate}].

The parameters of the drive are found by equating the total evolution matrix of the driven qubit to the matrix of the required operation, multiplied by the factor $e^{i\varphi}$ with an arbitrary $\varphi$, as it does not affect the dynamics of the system [see Eq.~\eqref{gate_equivalence}]:
\begin{equation}
	R_z(\phi_\text{pre}) \Xi_\text{Q} R_z(\phi_\text{after}) = e^{i\varphi} H.
	\label{drive_condition}
\end{equation}

Here we describe a \textit{simple algorithm for finding the optimal parameters of the drive} with four LZSM transitions $\mathcal{P}$, $\Phi_\mathrm{St}$, $A$, $\omega$, $t_\text{pre}$, $t_\text{after}$ that implements the Hadamard gate $H$.

The final upper energy-level occupation is the occupation probability of the exited state of the qubit $\ket{E_+}$
after applying the drive with four LZSM transitions to the qubit in the ground state $\ket{E_-}$, and is given by
\begin{equation}
	\mathcal{P}_\text{final} = \left\vert \Xi_{\text{Q}12} \right\vert^2.
	\label{p_final}
\end{equation}

\begin{figure}[t]
	\center{
		\includegraphics[width=1							\columnwidth]{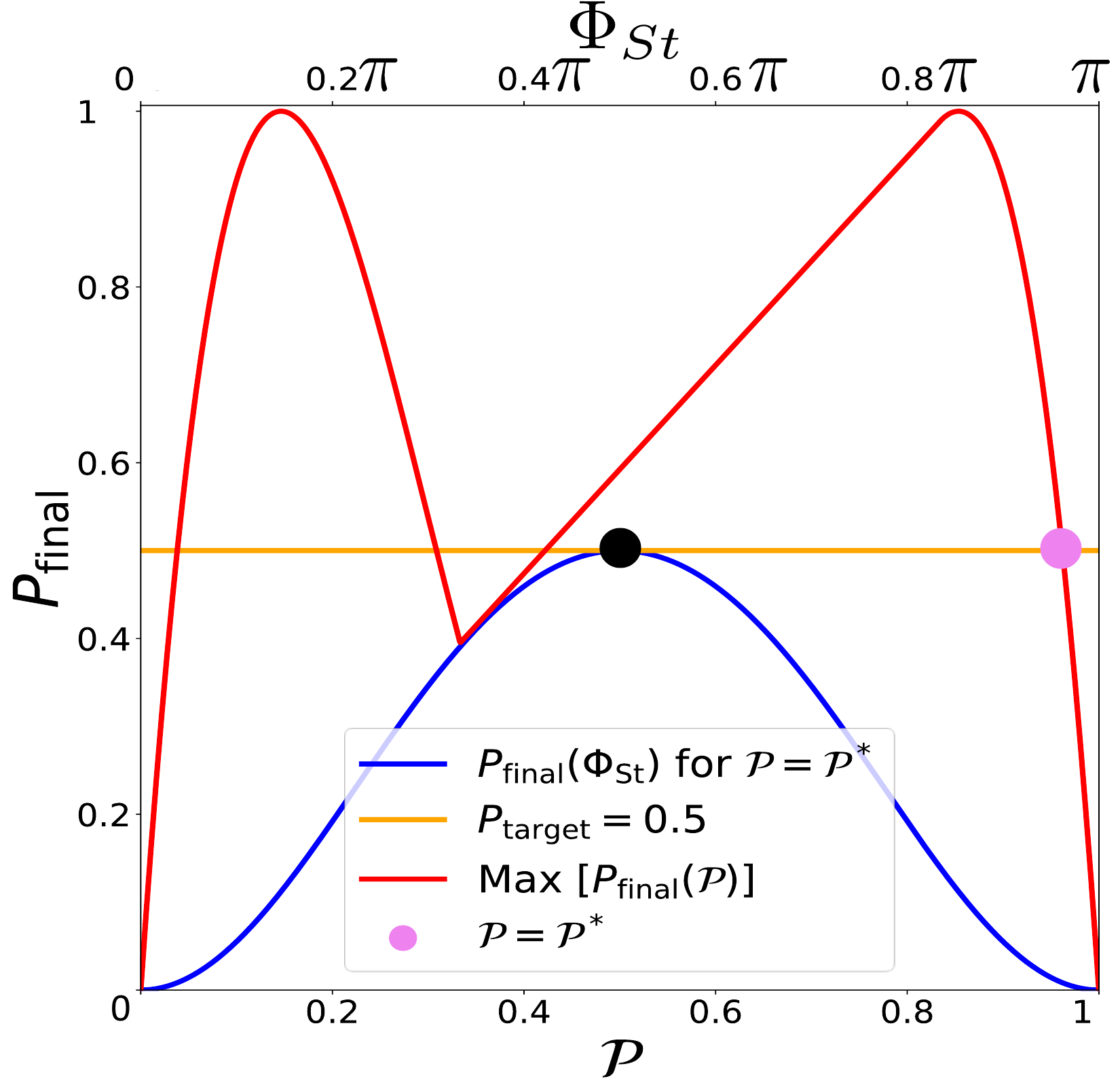}}
	\caption{Graphical demonstration of finding the values of a single LZSM transition probability $\mathcal{P}$ and St\"{u}ckelberg phase $\Phi_\text{St}$, for the implementation of the Hadamard gate 
		$H$ with four LZSM transitions. Red curve: maximum possible final occupation probability after four transitions depending on $\mathcal{P}$.
		Blue curve: dependence of the final occupation probability after four transitions on the St\"{u}ckelberg phase gain during single-transition $\Phi_\text{St}$.
		The orange horizontal line: level of $\mathcal{P}_\text{final} = \mathcal{P}_\text{target}$. }
	\label{Fig:MaxProbability changingQuadro transitions}
\end{figure}

$\bullet$ First, we find the possible values of a single LZSM transition probability $\mathcal{P}$ that provide the target final upper energy-level occupation probability after four transitions $\mathcal{P}_\text{final} = \mathcal{P}_\text{target}$.
These can be found as the crossings of the red curve and orange horizontal line in Fig.~\ref{Fig:MaxProbability changingQuadro transitions}. The red curve represents the maximum possible final upper energy-level occupation after four LZSM transitions $\mathcal{P}_\text{final}$ after varying through all possible values of $\Phi_\text{St}\in[0,\pi]$ using Eqs.~\eqref{sigma_q12} and~\eqref{p_final}.
The orange horizontal line represents the level of $\mathcal{P}_\text{final} = \mathcal{P}_\text{target}$.
For the $H$ operation, the target final upper energy-level occupation of the qubit $\mathcal{P}_\text{target} = 1/2$.
Larger values of $\mathcal{P}$ provide shorter transition durations and shorter gate durations, so the largest possible value of $\mathcal{P}$ is selected. For the $H$ operation realized with four LZSM transitions the largest possible value $\mathcal{P^*} \approx 0.962$.
For the $X$ operation with four LZSM transitions, $\mathcal{P^*}=(2+\sqrt{2})/4$ coincides with the LZSM transition probability for the $H$ operation, realized with two LZSM transitions, see Eq.~\eqref{H_two_passages_condition}.

$\bullet$ At the second step, we find the St\"{u}ckelberg phase $\Phi_\text{St}$ that provides the target final transition probability
$\mathcal{P}_\text{final} = \mathcal{P}_\text{target}$ given the obtained value of $\mathcal{P}$ using the blue curve in Fig.~\ref{Fig:MaxProbability changingQuadro transitions}.
For the $H$ operation the solution is given by
\begin{equation}
	\Phi_\text{St}=\frac{\pi}{2}+\pi n,
\end{equation}
where $n$ is an integer.
This is also a condition for the constructive interference between the LZSM transitions.

\begin{figure}[t]
	\center{
		\includegraphics[width=1	\columnwidth]{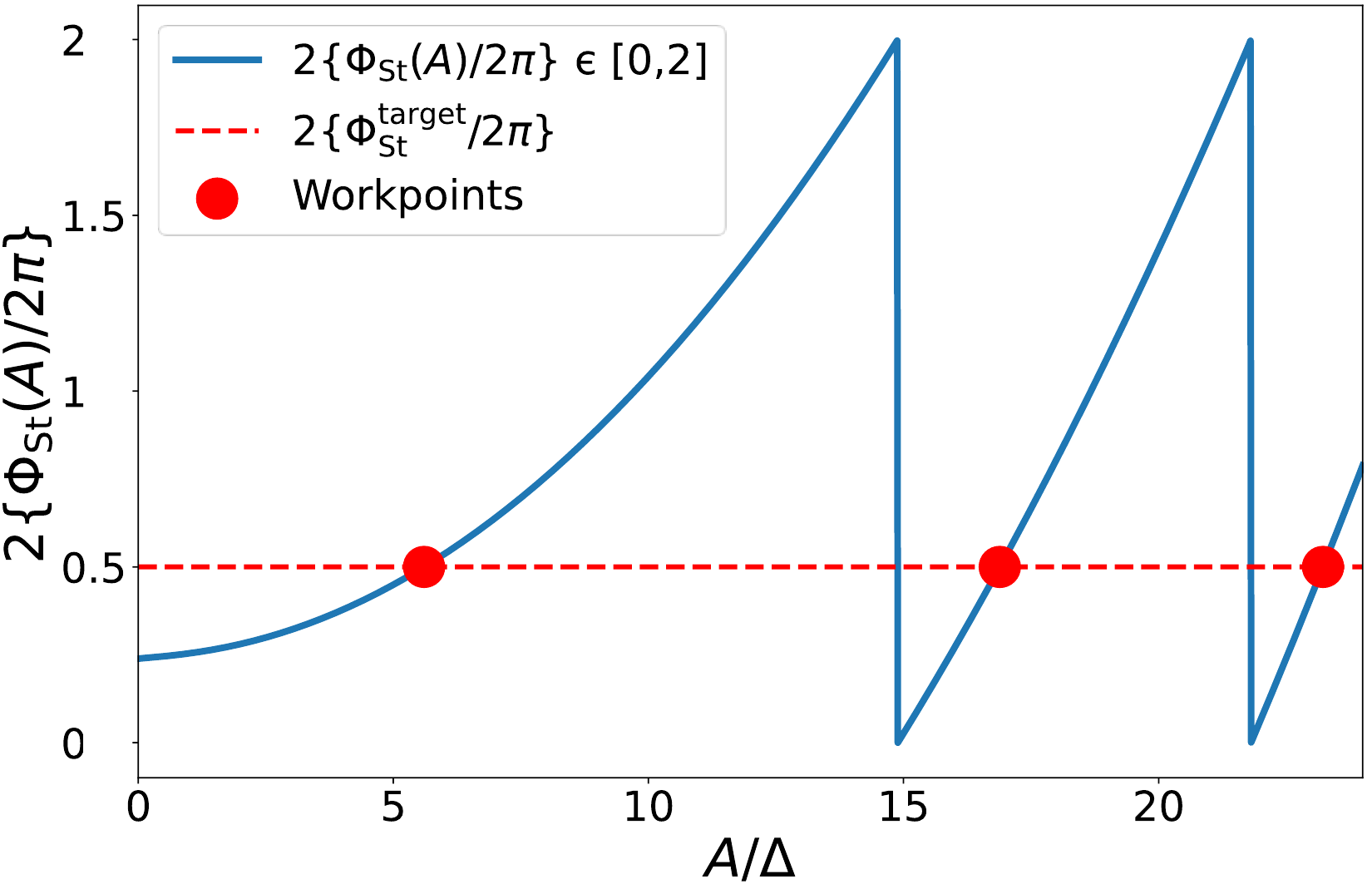}}
	\caption{Graphical procedure of finding possible values of the amplitude $A$ of the drive with four LZSM transitions for the implementation of the Hadamard gate $H$.
		The dependence of $2 \left\{\Phi_\text{St}/2\pi\right\}$ on the amplitude $A$ of the drive for $\mathcal{P}=\mathcal{P^*}$. Here, $\Phi_\text{St}$ is the St\"{u}ckelberg phase for the given value of the amplitude $A$; while the target $\Phi_\text{St}$ is the required value of the St\"{u}ckelberg phase of the drive, found in the previous step, and \{\} is the fractional part. Red dots represent the possible values of the amplitude $A$ of the drive with four LZSM transitions that implement the Hadamard operation $H$. }
	\label{Fig:AWPhase calc}
\end{figure}

$\bullet$ After finding $\mathcal{P}$ and $\Phi_\text{St}$, at the third step, we find the amplitude $A$ and frequency $\omega$ of the signal ~\eqref{LZSMdrive}.
Equation~\eqref{AwLZSMConnection} defines the relation between the frequency $\omega$ and the amplitude $A$ for a certain $\mathcal{P}$.
Using Eqs. ~\eqref{AwLZSMConnection}, ~\eqref{PhiSt}, ~\eqref{StokesPhase}, ~\eqref{delta} and ~\eqref{zeta}, we build Fig.~\ref{Fig:AWPhase calc} and find the possible values of the amplitude $A$ of the signal that provides the required value of the St\"{u}ckelberg phase $\Phi_\text{St target}$, found in the previous step for the previously found $\mathcal{P}=\mathcal{P^*}$.

$\bullet$ Finally, at the fourth step we determine the required idling durations before and after the main part of the drive with LZSM transitions,  $t_\text{pre}$ and $t_\text{after}$. Substitution of the obtained evolution matrix of the main part of the drive $\Xi_\text{Q}$ to Eq.~\eqref{drive_condition} allows to find the required accumulated phases $\phi_\text{pre}$ and $\phi_\text{after}$. Then using Eqs.~\eqref{LarmorFrequency} and ~\eqref{t_phase_gate} we determine the durations $t_\text{pre}$ and $t_\text{after}$. In the example considered here, the accumulated phases $\phi_\text{pre} = \phi_\text{after} = \pi/2$, and the durations $t_\text{pre}$ and $t_\text{after}$ depend on the choice of the amplitude $A$ in the previous step of the algorithm. 

This algorithm of finding the parameters of the drive  ~\eqref{LZSMdrive} with an even number of LZSM transitions for the implementation of any single-qubit operation can be summarized as follows:
\begin{itemize}
	\item Find the probability of a single LZSM transition $\mathcal{P}$ which provides the desired final upper-level occupation probability $\mathcal{P}_\text{final} = \mathcal{P}_\text{target}$. See the red curve cross with the orange horizontal line in Fig.~\ref{Fig:MaxProbability changingQuadro transitions}. 
	\item Find the required St\"{u}ckelberg phase $\Phi_\text{St}$. See the blue curve in Fig.~\ref{Fig:MaxProbability changingQuadro transitions}.
	\item Find the combination of the amplitude $A$ and the frequency $\omega$ that provides the required values of $\mathcal{P}$ and $\Phi_\text{St}$. See Fig.~\ref{Fig:AWPhase calc}.
	\item Determine the idling times before and after the main part of the drive with LZSM transitions, $t_\text{pre}$ and $t_\text{after}$.
\end{itemize}

This algorithm allows to find the optimal parameters of the drive with an arbitrary even number of LZSM transitions. Here we consider only a two-level quantum system, but real quantum systems are usually multilevel. The LZSM transitions during the passage of nearest level anti-crossings with different energy levels will influence the dynamics, so it is important to limit the amplitude of the drive, so that the next nearest anti-crossings are not reached.

\subsection{Fidelity}
The relaxation and dephasing are not considered in this paper. Thus the infidelities of the gates arise because the theories of RWA and AIM that are used to obtain the parameters of the driving signals are approximate. The infidelities due to numerical solution errors are negligible in comparison with infidelities due to approximations in the theories.

The fidelities are found using quantum tomography \cite{Vitanov2020}, which consists in applying the gate for many different initial states, which span the Hilbert space, and then calculating the average fidelity between the obtained states and the target state using \cite{Jozsa1994} 
\begin{equation}
	F(\rho,\rho_\text{t})=\left(\text{tr}\sqrt{\sqrt{\rho} \rho_\text{t}\sqrt{\rho}}\right)^2.
\end{equation}
Here, $\rho$ is the density matrix obtained by numerical simulation of the qubit dynamics by solving the Liouville-von Neumann equation Eq.~\eqref{VonNeumann}, and  $\rho_\text{t}=U\rho_\text{in}U^\dagger$ is the target state, obtained by applying the gate operator $U$ to the initial density matrix $\rho_\text{in}$.
Then we calculate the averaged fidelity for different equidistant initial conditions on the Bloch sphere $\bar{F}=\sum_{n=1}^N F(\rho_n, \rho_{\text{t}_n})/N$.
\begin{figure}[h]
	\center{
		\includegraphics[width=1							\columnwidth]{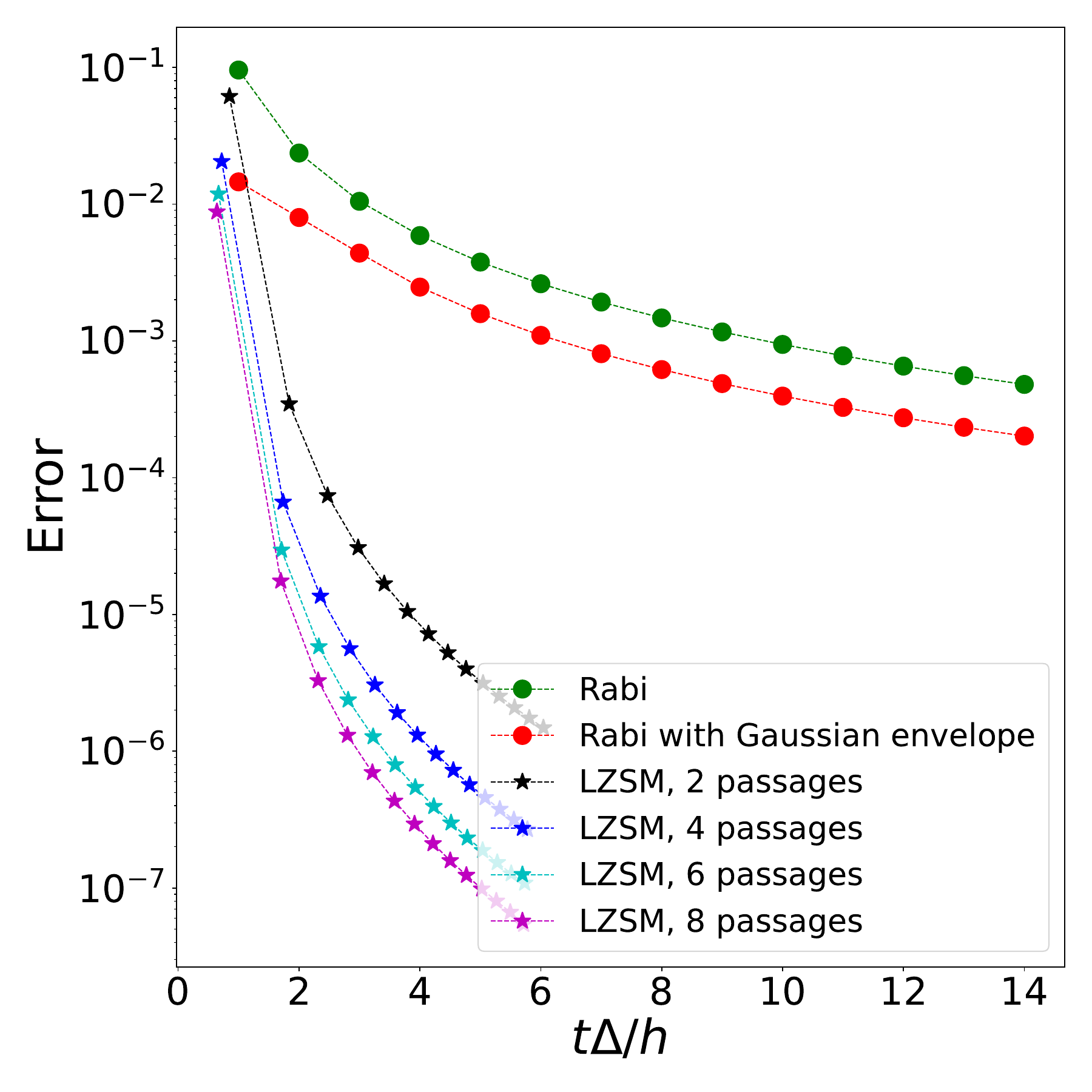}}
	\caption{Comparison of error rates for Rabi and LZSM approaches for single-qubit gate implementations. The error rates  (1-$\bar{F}$) for Rabi and LZSM approaches for single-qubit $Y$ gate implementation calculated for different gate durations. For the Rabi implementation the gate duration equals an integer number of periods of a resonant driving (from 1 to 14). The Gaussian envelope is cut at $G=2.7$, see Eq.~\eqref{G_Gaussian}.  For the LZSM implementation, the gate durations are built for different workpoints using  Fig.~\ref{Fig:AWPhase calc} and the analogous figures built for higher number of LZSM transitions. The error for the LZSM approach decreases much faster than for the Rabi approach.}
	\label{Fig:Fidelity}
\end{figure}
To better compare the difference between Rabi and LZSM approaches we will use the error rate $D=1-\bar{F}$.

The LZSM probability formula $\mathcal{P}=\exp{(-2\pi\delta)}$ is derived for a linear drive with an infinite time, $\varepsilon (t) = vt$, $t \in (-\infty, \infty)$, leading to an infinite amplitude of the driving signal. Thus, for the considered non-linear drive $\varepsilon(t)=-A\cos(\omega t)$ with the finite amplitude $A$, the fidelity of the LZSM gate increases with the amplitude of the drive $A$.
Considering Eq.~\eqref{AwLZSMConnection}, the amplitude of the drive is proportional to the duration of the gate, $A \sim 1/\omega \sim T$. So the fidelity of the LZSM gate increases with its duration, the gate error $D$ decreases,  and a satisfactory balance between the fidelity and speed of the gate should be found. 
Figure~\ref{Fig:Fidelity} illustrates that the gate error rate $D$ using the LZSM implementation decreases much faster with time than using the usual Rabi approach.

An alternative method to determine the gates fidelity is the Randomized benchmarking \cite{Jones2021}, which considers how the fidelity decreases with increasing the number of applied operations. Here we only used the quantum tomography method, as a simpler one for numerical calculations.

In experiments, there are methods of improving the gate fidelity based on the back-response loop, also know as quantum control or robust control, for example gradient ascent or Krotov's method \cite{Kuzmanovic2023,Araki2023,Huang2014,Li2022}.

\section{Two-qubit gates}
\label{Sec:Two-qubit gates}
\subsection{Hamiltonian}
\begin{figure}[t]
	\centering{
		\includegraphics[width=1 \columnwidth]{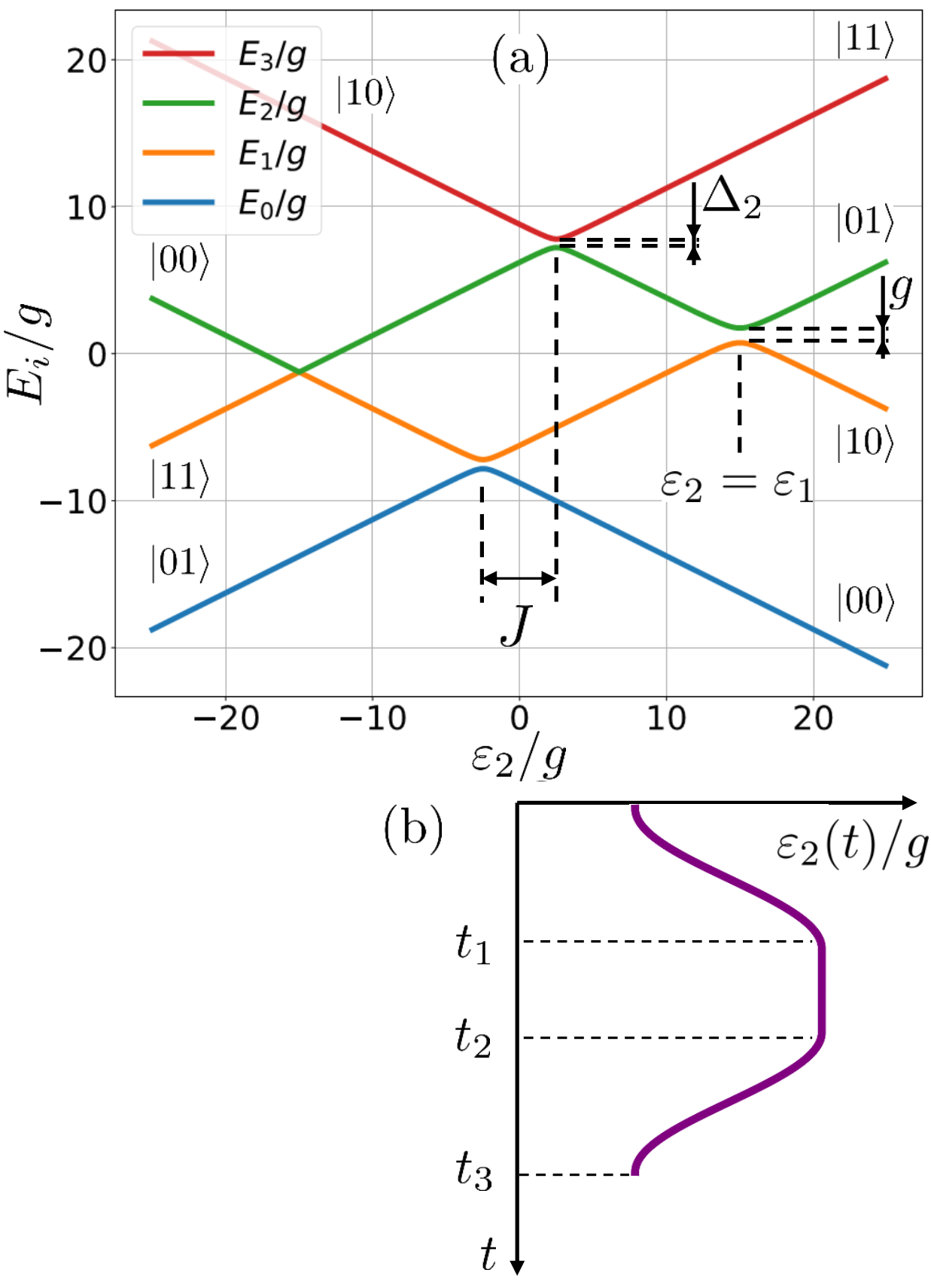}}
	\caption{ 
		(a) Adiabatic energy levels of a two-qubit system with Hamiltonian \eqref{Hamiltonian} as a function of the driving parameter $\varepsilon_2$.
		In the regions far from the level anti-crossings, the energies of the adiabatic levels $\ket{E_i}$ asymptotically coincide with the energies of the diabatic levels $\ket{mn}$.
		The parameters of the Hamiltonian used here are: $\Delta_1/g=0.6$,  $\Delta_2/g=0.6$, $\varepsilon_1/g=15$, and $J/g=5$.  
		(b) Time dependence of the driving parameter $\varepsilon_2(t)$ for the iSWAP gate implementation.}
	\label{Fig:energy_diab_adiab_with_signal_cnot_normalized}
\end{figure}

Now we consider a Hamiltonian of two coupled qubits \cite{Krantz2019}
\begin{equation}
	\begin{gathered}
		H = -\frac{1}{2} \sum \limits_{i=1,2} ( \Delta_i \sigma^{(i)}_x + \varepsilon_i (t) \sigma^{(i)}_z ) -\\-  \frac{g}{4} ( \sigma^{(1)}_x \sigma^{(2)}_x + \sigma^{(1)}_y \sigma^{(2)}_y )  - \frac{J}{4} \sigma^{(1)}_z \sigma^{(2)}_z,
	\end{gathered}
	\label{Hamiltonian}
\end{equation}
with the external drive of the second qubit which results in the driving $\varepsilon_2(t)$.
The energy levels of this Hamiltonian normalized to the coupling strength $g$ are shown in Fig.~\ref{Fig:energy_diab_adiab_with_signal_cnot_normalized}(a).

Although other choices for the interaction part of the Hamiltonian are possible, we will now consider a transverse coupling with $XY$-interaction
\begin{equation}
	H^{XY}_{\text{int}} = -\frac{g}{4} ( \sigma^{(1)}_x \sigma^{(2)}_x + \sigma^{(1)}_y \sigma^{(2)}_y ),
\end{equation}
resulting in a splitting between $E_{1}$ and $E_{2}$ adiabatic energy levels  versus $g$ at the crossing of $\ket{01}$ and $\ket{10}$ diabatic energy levels, and longitudial couplings with $ZZ$-interaction term
\begin{equation}
	H^{ZZ}_{\text{int}} = -\frac{J}{4}  \sigma^{(1)}_z \sigma^{(2)}_z,
\end{equation}
resulting in a shift between the  $(E_{0}-E_{1})$ and $(E_{2}-E_{3})$ adiabatic energy-level anti-crossings on the value of $J$ [see Fig.~\ref{Fig:energy_diab_adiab_with_signal_cnot_normalized}(a)].
The $JJ$- or Heisenberg interaction is the particular case when both terms are present and $J=g$.

The difficulty of generating a particular operation depends on the available coupling terms. On the other hand, for each type of coupling there are two-qubit gates which can be implemented in a straightforward manner \cite{Schuch2003}.

\subsection{iSWAP gate}

One of the simplest natural two-qubit operations when the $XY$-type of coupling is present, is an iSWAP gate
\begin{equation}
	\begin{gathered}
		\text{iSWAP}	=\begin{pmatrix} 
			1 & 0 & 0 & 0 \\
			0 & 0 & i & 0 \\
			0 & i & 0 & 0 \\
			0 & 0 & 0 & 1  \\  \end{pmatrix}.
	\end{gathered}	\label{iswap}
\end{equation}
Its LZSM implementation should involve passages of the anti-crossing between the adiabatic levels $E_{1}$ and $E_{2}$, located at $\varepsilon_2 = \varepsilon_1$.
For simplicity, here we demonstrate an LZSM realization of the iSWAP gate for the Hamiltonian \eqref{Hamiltonian} with only $XY$-interaction term, when $J=0$ and the  $(E_{0}-E_{1})$ and $(E_{2}-E_{3})$ anti-crossings are both located at $\varepsilon_2=0$ [see Fig.~\ref{Fig:energy_diab_adiab_with_signal_cnot_normalized}(a)].

\begin{figure}[t]
	\centering{%
		\includegraphics[width=1 \columnwidth]{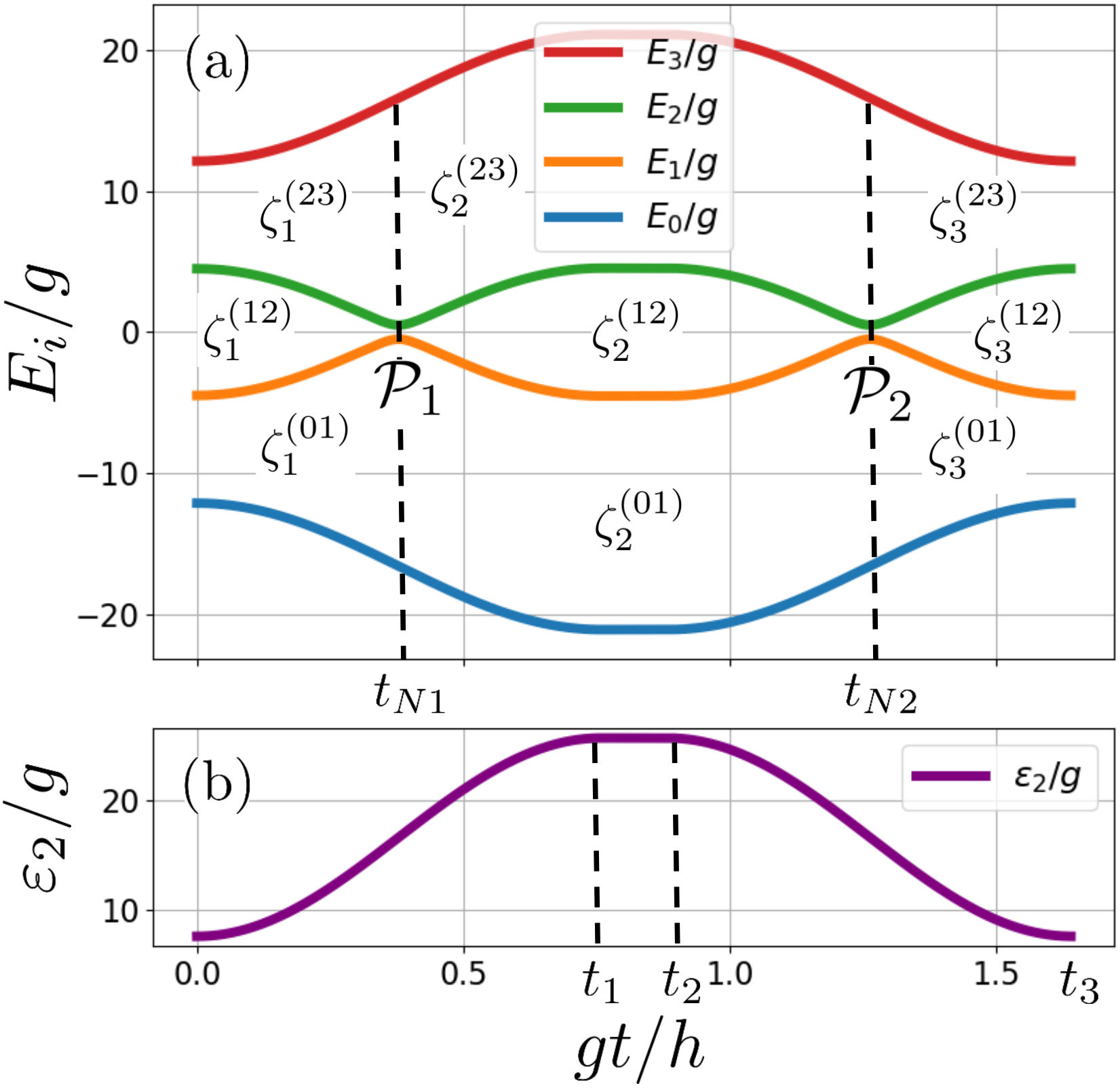}}
	\caption{Dependence of the adiabatic energy levels $E_i$, (a), and the driving energy detuning $\varepsilon_2$, (b), versus time for the drive \eqref{iswap_signal} for implementing the iSWAP gate. For each diabatic transition $N_i$ during the level anti-crossing at time $t_{N(i)}$, the value of the LZSM probability $\mathcal{P}_i$ is introduced. For each adiabatic evolution interval, the phase gains $\zeta^{(ij)}_k$ between various adiabatic levels $E_i$ and $E_j$ can be represented as areas between the levels.
	}
	\label{Fig:iswap_energy_dynamics}
\end{figure}

As in the case of the $X$ gate, it is impossible to implement an LZSM transition with an arbitrary $\mathcal{P}$ with high fidelity by only one passage, so at least two passages are required.
Thus, we consider a drive $\varepsilon_2(t)$ with the following form [see Figs.~\ref{Fig:energy_diab_adiab_with_signal_cnot_normalized}(b) and~\ref{Fig:iswap_energy_dynamics}(b)]:
\begin{equation}
	\varepsilon_2(t) =
	\begin{cases}
		\varepsilon_1  - A \cos \omega t, \  \ \  \ \  \ \  \ \  \ \ \ \  \ \ \  \ \  0 < t < t_1,	\\
		\varepsilon_1 + A  ,   \ \  \ \  \ \  \ \  \ \  \ \  \ \  \ \  \ \  \ \  \ \  \ \ \ \  t_1 < t < t_2,	\\
		\varepsilon_1 + A \cos \omega (t-\frac{T_\text{c}}{2} - T_1),  \ \  t_2 < t < t_3,
	\end{cases}
	\label{iswap_signal}
\end{equation}
where
\begin{equation}
	\begin{aligned}
		& t_1 = \frac{T_\text{c}}{2},		\\
		& t_2 = \frac{T_\text{c}}{2} + T_1, 	\\
		& t_3 = T_\text{c} + T_1. 
	\end{aligned}
	\notag
\end{equation}
It consists of two half-periods of cosine with period $T_\text{c} ={2 \pi}/{\omega}$ and amplitude $A$, separated in the middle by an idling period with time $T_1$.
For the simpler form of a signal without the idling we would obtain a system of three equations on the signal parameters with only two parameters present, $A$ and $T_\text{c}$. So an additional degree of freedom, like an idling time $T_1$, would be needed.

As in the case of a single qubit, we build the dependence of the adiabatic energy levels on time in Fig.~\ref{Fig:iswap_energy_dynamics}(a), introduce all values of the transition probabilities $\mathcal{P}_i$ for each diabatic transition $N_i$, and define the phase gains $\zeta^{(ij)}_k$ between adiabatic levels $E_i$ and $E_j$ for various periods of the adiabatic evolution as
\begin{equation}
	\zeta^{(ij)}_k = \frac{1}{2 \hbar} \int_{t_{N(k-1)}}^{t_{N(k)}} 
	\left[ E_j(t) - E_i(t) \right] dt,
	\label{zeta_multi}
\end{equation}
where $t_{N0} = 0$, and $t_{N3} = t_3$.

Generally, for a multi-level quantum system, the matrix of the \textit{diabatic} (LZSM) transition between the adiabatic energy levels $\ket{E_i}$ and $\ket{E_j}$ ($i<j$) with the LZSM probability $\mathcal{P}$ in the adiabatic basis is defined as
\begin{equation}
	\begin{gathered}
		N = R e^{i\phi_\text{S}} \ket{E_i}\bra{E_i}
		+ R e^{-i\phi_\text{S}} \ket{E_j}\bra{E_j} + \\
		+ \alpha T \ket{E_i}\bra{E_j}
		- \alpha T \ket{E_j}\bra{E_i}
		+ \sum_{k \ne i,j} \ket{E_k}\bra{E_k},
	\end{gathered}
\end{equation}
where the transition and reflection coefficients, $T$ and $R$, and the Stokes phase $\phi_\text{S}$ are determined  by the LZSM probability $\mathcal{P}$, see Eqs.~\eqref{tran_refl_coef} and~\eqref{StokesPhase}.
The coefficient $\alpha$ depends on the direction of the passage of the adiabatic energy levels anti-crossing.
Far from the anti-crossing region, the energies of the adiabatic states $\ket{E_i}$ and $\ket{E_j}$ asymptotically approach the energies of some diabatic states $\ket{m}$ and $\ket{n}$, where $m<n$. Here, we assume that the diabatic basis $\{ ..., \ket{m},..., \ket{n},...  \}$ is the one, in which the Hamiltonian is defined. If before the passage of the adiabatic energy levels anti-crossing energy of the lower adiabatic level $\ket{E_i}$ is asymptotically close to the energy of the diabatic level with the lower sequence number $\ket{m}$, then $\alpha=1$. If before the passage $\ket{E_i}$ is asymptotically close to $\ket{n}$, then $\alpha=-1$.

For the considered Hamiltonian \eqref{Hamiltonian}, defined in the diabatic basis $\{ \ket{00}, \ket{01}, \ket{10}, \ket{11} \}$,
and the drive \eqref{iswap_signal}, the
matrices of the diabatic transitions are defined by
\begin{equation}
	\begin{gathered}
		N_k	=\begin{pmatrix}
			1 & 0 & 0 & 0 \\
			0 & R_k e^{i  \phi_{\text{S}(k)}} &  \alpha_k T_k & 0 \\
			0 & - \alpha_k T_k & R_k e^{-i  \phi_{\text{S}(k)}} & 0	\\
			0 & 0 & 0 & 1  \end{pmatrix},
	\end{gathered}	
\end{equation}
where $k=1,2$, $\alpha_1 = 1$, $\alpha_2 = -1$.
Here, $T_i$, $R_i$, $\phi_{\text{S}(i)}$ are the transition, reflection coefficients, and the Stokes phase for the diabatic transition $N_i$.

The matrix for the \textit{adiabatic} evolution $U_n$ for the interval of evolution $n=1,2,3$ is diagonal with components
\begin{eqnarray}
	U_{(n)00}&=&e^{i (\zeta^{(01)}_n + \zeta^{(12)}_n + \zeta^{(23)}_n)},\notag\\
	U_{(n)11}&=&e^{i ( -\zeta^{(01)}_n + \zeta^{(12)}_n + \zeta^{(23)}_n)},\\\notag
	U_{(n)22}&=&e^{i (- \zeta^{(01)}_n - \zeta^{(12)}_n + \zeta^{(23)}_n )},\\\notag
	U_{(n)33}&=&e^{ -i  (\zeta^{(01)}_n + \zeta^{(12)}_n + \zeta^{(23)}_n)}.
	\label{U_adiabatic_evolution}
\end{eqnarray}
In the case of a $n$-level quantum system, the matrix
of the adiabatic evolution is diagonal with components
\begin{equation}
	\begin{gathered}
		U_{kk} = \exp{ \left\{ i  \sum_{j=0}^{n-2} \beta_{jk} \zeta^{(j,j+1)} \right\} },	\\
		\beta_{jk} = 2\theta(j-k)-1,
	\end{gathered}
\end{equation}
where $k=0,1,...,n-1$, and $\theta$ is the Heaviside step function. The rule of the sign $\beta_{jk}$ determination can be summarized as the following:
if the area of the phase accumulation, corresponding to $\zeta^{(j,j+1)}$ term is below the adiabatic energy level $E_k$, then $\beta_{jk}=-1$; if above it, then $\beta_{jk}=1$.

The evolution matrix for the whole period becomes
\begin{equation}
	\begin{gathered}
		\Xi = U_3 N_2 U_2 N_1 U_1.
	\end{gathered}
	\label{total_evolution_matrix}
\end{equation}
After simplifying by
\begin{equation}
	\begin{gathered}
		\zeta^{(01)} = \zeta^{(01)}_1 + \zeta^{(01)}_2 + \zeta^{(01)}_3,	\\
		\zeta^{(23)} =	\zeta^{(23)}_1 + \zeta^{(23)}_2 + \zeta^{(23)}_3,
	\end{gathered}
\end{equation}
taking the common phase $e^{i \xi}$ out of brackets and neglecting it (as the common phase of the wave function after the logic gate is irrelevant), we obtain the evolution matrix
\begin{equation}
	\begin{gathered}
		\Xi = \begin{pmatrix} 
			1 & 0 & 0 & 0 \\
			0 & U_{11} & U_{12} & 0 \\
			0 & U_{21} & U_{22} & 0 \\
			0 & 0 & 0 & U_{33}    \end{pmatrix},
	\end{gathered}\label{iswap_evolution}
\end{equation}
that depends on the values $\mathcal{P}_1, \mathcal{P}_2, \zeta^{(01)}, \zeta^{(12)}_i, \zeta^{(23)}$.
Equating it to the matrix of a required two-qubit iSWAP gate allows to determine the parameters of the external signal which implements this gate:
\begin{equation}
	\begin{aligned}
		&\mathcal{P}_1+\mathcal{P}_2  = 1,	\\
		&\phi_\text{S1} + \phi_\text{S2} + 2\zeta^{(12)}_2  = \pi + 2 \pi n_1,	\\
		&\phi_\text{S1} +	2(\zeta^{(01)} + \zeta^{(12)}_1 + \zeta^{(12)}_2 ) = \frac{\pi}{2} + 2 \pi n_2, \\
		&\phi_\text{S2} + 2(\zeta^{(01)} + \zeta^{(12)}_2 + \zeta^{(12)}_3 ) = \frac{\pi}{2} + 2 \pi n_3.
	\end{aligned}\label{iswap_p1_p2}
\end{equation}

\begin{figure}[t]
	\centering{%
		\includegraphics[width=1	\columnwidth]{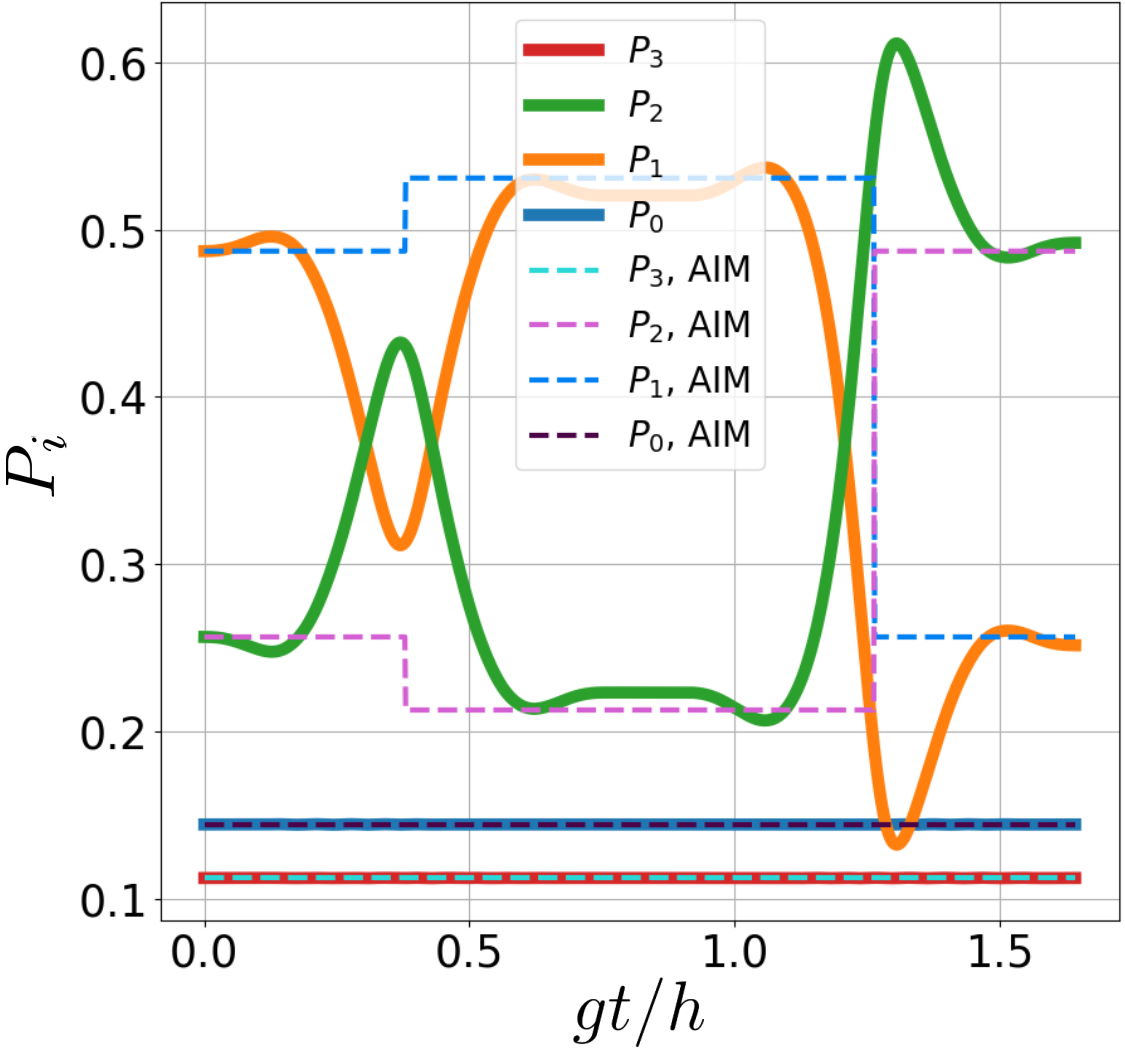}}
	\caption{ The dynamics of the iSWAP gate implemented with two LZSM transitions. The occupation probabilities $P_i$ of each adiabatic level $E_i$ as function of time are obtained by two methods: numerical solutions of the Liouville-von Neumann equation and by the adiabatic-impulse model.
		The iSWAP gate results in the swap of occupation probabilities between the $E_1$ and $E_2$ adiabatic energy levels.} 
	\label{Fig:iSWAP_dynamics}
\end{figure}

For the considered signal \eqref{iswap_signal} with $\mathcal{P}_1 = \mathcal{P}_2 = \mathcal{P}$, resulting in $\zeta^{(12)}_1 = \zeta^{(12)}_3$, and in the case when only the $XY$-interaction is present ($J=0$), resulting in $\zeta^{(01)} = \zeta^{(23)}$, the conditions simplify to
\begin{equation}
	\begin{aligned}
		&\mathcal{P} = \frac{1}{2},	\\
		&\phi_\text{S} + \zeta^{(12)}_2  = \frac{\pi}{2} +  \pi n_1,	\\
		&\phi_\text{S} +	2(\zeta^{(01)} + \zeta^{(12)}_1 + \zeta^{(12)}_2 ) = \frac{\pi}{2} + 2 \pi n_2.
	\end{aligned}\label{iswap_conditions}
\end{equation}

The first equation results in a linear dependence between $A$ and $T_\text{c}$:
\begin{equation}
	\begin{gathered}
		A = \frac{g^2}{4 \hbar \ln{2}} T_\text{c}.
	\end{gathered}	
\end{equation}
Then we numerically solve two other equations on two parameters of the signal $T_\text{c}$ and $T_1$.
In Fig.~\ref{Fig:iSWAP_dynamics} we demonstrate the dynamics of the iSWAP gate for a particular solution with $g T_\text{c}/h = 1.517$, $g T_1/h = 0.125$, and $A/g=3.438$ for the Hamiltonian \eqref{Hamiltonian} with the parameters $\Delta_1/g=0.3$,  $\Delta_2/g=1$, $\varepsilon_1/g=16.6$, $J=0$; and compare the approximate solution obtained by the adiabatic-impulse model with the numerical solution of the Schr\"{o}dinger equation. The same parameters of the Hamiltonian and the drive were used for Fig.~\ref{Fig:iswap_energy_dynamics}, with the exception of the larger amplitude $A/g=9$.

\section{Conclusion}
\label{Sec:Conclusion}

We further developed the paradigm of the alternative quantum logic gates, based on the LZSM transitions.
We demonstrated how the adiabatic-impulse model can be used for implementing single- and two-qubit gates, demonstrated how to increase the gate speed, and the technique of finding the balance between speed and fidelity of the gates.
We also demonstrated the comparison of the theoretical error rate for conventional Rabi gates and alternative LZSM gates for various logic gate durations.

The adiabatic-impulse model is applicable for any quantum multi-level systems with two conditions. Firstly, it works well for a large drive amplitude, $A > \Delta$. In terms of the requirements for a quantum system, this means that for the considered level anti-crossing, its minimal energy splitting $\Delta$ should be much less than the distance to the nearest level anti-crossings. Secondly, the time between the LZSM transitions should be larger than the time needed for the transition process. This condition limits the maximal frequency of the driving signal in the multi-passage implementation, and the minimal gate duration, respectively.

An arbitrary single-qubit quantum logic gate can be performed with only two LZSM transitions. However, the considered option of gate implementation with multiple LZSM transitions provide a better combination of gate duration and fidelity. We demonstrated the technique of implementing an arbitrary single-qubit logic gate with any number of the LZSM transitions.

For the multi-level quantum systems the considered general method of implementing quantum logic gates with LZSM transitions is the following: choose the shape of the driving signal so that it passes the required level anti-crossings for a given gate. For the considered signal compute the dependence of the adiabatic energy levels of the system on time. Introduce the transition probabilities $\mathcal{P}_i$ for each diabatic transition and phase gains $\zeta^{(ij)}_k$ between all pairs of successive adiabatic levels $E_i$ and $E_j$ for each period of adiabatic evolution. Using them, compose all matrices of the diabatic transition $N_i$ and adiabatic evolution $U_i$, multiply them, and obtain the total evolution matrix.
Equating it to the matrix of the required quantum logic gate multiplied by an arbitrary phase term $e^{i \varphi}$ allows to determine the required parameters of the driving signal,
that implements this logic gate.

Note added. After this work was completed, we became aware of a recent relevant preprint \cite{Caceres2023}.

\begin{acknowledgments}
	The research of A.I.R, O.V.I., and S.N.S. is sponsored by the Army Research Office under Grant No. W911NF-20-1-0261.
	A.I.R. and O.V.I. were supported by the RIKEN International Program Associates (IPA).
	S.N.S. is supported in part by the Office of Naval Research Global, Grant No. N62909-23-1-2088.
	F.N. is supported in part by: Nippon Telegraph and Telephone Corporation (NTT) Research, the Japan Science and Technology Agency (JST) [via the Quantum Leap Flagship Program (Q-LEAP), and the Moonshot R\&D Grant Number JPMJMS2061], Office of Naval Research (ONR), and the Asian Office of Aerospace Research and Development (AOARD) (via Grant No. FA2386-20-1-4069).
\end{acknowledgments}

\appendix


\section{iSWAP-like gates. SWAP, $\sqrt{\text{SWAP}}$, $\sqrt{\text{iSWAP}}$}
\label{Sec:AppendixA}

The evolution matrix \eqref{iswap_evolution} can also implement the SWAP, $\sqrt{\text{SWAP}}$, $\sqrt{\text{iSWAP}}$, and other iSWAP-like gates.
The system of equations \eqref{iswap_p1_p2} written for the SWAP gate is not compatible. Thus, the SWAP gate cannot be implemented by only two passages of the $(E_{1}-E_{2})$ adiabatic energy-level anti-crossing for the Hamiltonian \eqref{Hamiltonian} with only $XY$-coupling. It can, however, be implemented when $\zeta^{(01)} \ne \zeta^{(23)}$, in case both $XY$- and $ZZ$-interactions are present, as in Fig.~\ref{Fig:energy_diab_adiab_with_signal_cnot_normalized}(a).
The corresponding conditions are written as
\begin{equation}
	\begin{aligned}
		&\mathcal{P}_1+\mathcal{P}_2  = 1,	\\
		&\phi_\text{S1} + \phi_\text{S2} + 2\zeta^{(12)}_2  = \pi + 2 \pi n_1,	\\
		&\phi_\text{S1} +	2(\zeta^{(01)} + \zeta^{(12)}_1 + \zeta^{(12)}_2 ) = (1+\lambda)\frac{\pi}{2} + 2 \pi n_2, \\
		&\phi_\text{S2} + 2(\zeta^{(01)} + \zeta^{(12)}_2 + \zeta^{(12)}_3 ) = (1+\lambda)\frac{\pi}{2} + 2 \pi n_3,	\\
		&\zeta^{(01)} +  \zeta^{(12)}_1 + \zeta^{(12)}_2 + \zeta^{(12)}_3 + \zeta^{(23)} = \pi n_4,
	\end{aligned}	
\end{equation}
where $\lambda=1$ for the SWAP gate and $\lambda=0$ for the iSWAP gate.

The  $\sqrt{\text{SWAP}}$ and  $\sqrt{\text{iSWAP}}$ gates do not provide a full swap of energy-level occupation probabilities between two levels when $\mathcal{P}=1$; so they could be implemented by a single passage of the $(E_{1}-E_{2})$ level anti-crossing.

\section{CNOT-like gates. CNOT, CZ, CPHASE}
\label{Sec:AppendixB}

$ZZ$-type couplings allow to implement CNOT, CZ, CPHASE gates. Their LZSM implementations should involve passages of the anti-crossing between adiabatic energy levels $E_{2}$ and $E_{3}$ at $\varepsilon_2=J/2$ [see Fig.~\ref{Fig:energy_diab_adiab_with_signal_cnot_normalized}(a)].
Here we demonstrate an LZSM realization of the CNOT gate for the Hamiltonian \eqref{Hamiltonian} with both $XY$- and $ZZ$-couplings, although only $ZZ$- is required.

\begin{figure}[t]
	\centering{
		\includegraphics[width=1 \columnwidth]{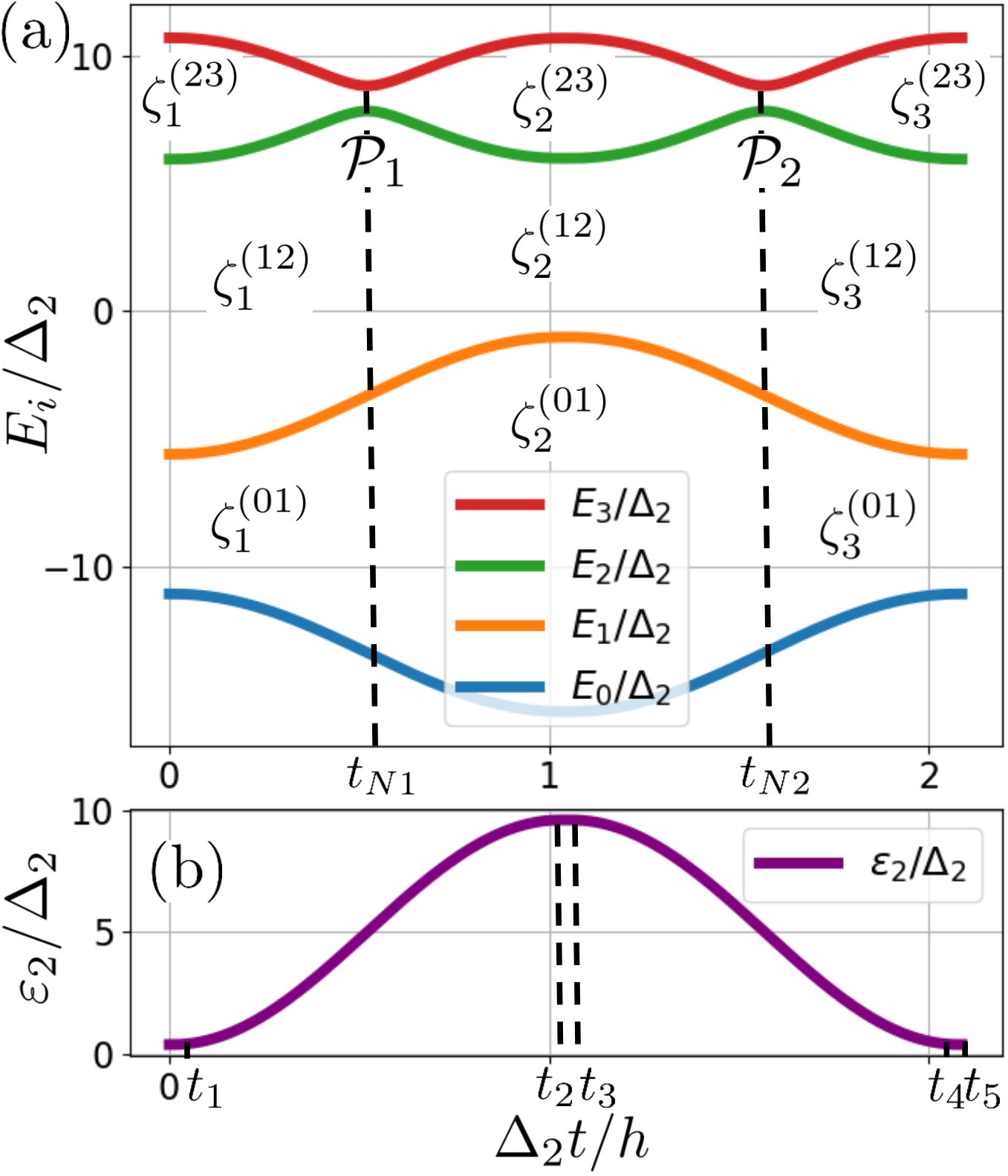}}
	\caption{ Dependence of the adiabatic energy levels $E_i$, (a), and the driving parameter, energy detuning $\varepsilon_2$, (b), versus time for the drive \eqref{cnot_signal} for implementing the CNOT gate. For each diabatic transition $N_i$ during the level anti-crossing at time $t_{N(i)}$, the value of the LZSM probability $\mathcal{P}_i$ is introduced. For each adiabatic evolution interval, the phase gains $\zeta^{(ij)}_k$ between various adiabatic levels $E_i$ and $E_j$, represented as areas between the levels.
	}
	\label{Fig:cnot_energy_dynamics}
\end{figure}

As for the $X$ gate, it is impossible to implement an LZSM transition with an arbitrary $\mathcal{P}$ with high fidelity by only one passage; so at least two passages are required.
We now consider a drive $\varepsilon_2(t)$ in the following form
[see Fig.~\ref{Fig:cnot_energy_dynamics}(b)]
\begin{equation}
	\begin{gathered}
		\varepsilon_2(t) =
		\begin{cases}
			J/2-A, \ \ \ \ \ \ \ \ \ \ \ \ \ \ \ \ \ \ \ \ \ \ \ \ \ \ \ \ \ \ \ \ \   0<t<t_1,	\\
			J/2 - A \cos \omega (t-T_1), \ \ \ \ \ \ \ \ \ \  \ \ \ \ \ \  t_1 < t < t_2,	\\
			J/2 + A, \ \ \ \ \ \ \ \ \ \ \ \ \ \ \ \ \ \ \ \ \ \ \ \ \ \ \ \ \ \ \ \ \  t_2 < t < t_3,	\\
			J/2 + A \cos \omega (t-T_1 - \frac{T_\text{c}}{2} - T_2),	\	\			t_3 < t < t_4,	\\
			J/2-A, \ \ \ \ \ \ \ \ \ \ \ \ \ \ \ \ \ \ \ \ \ \ \ \ \ \ \ \ \ \ \ \ \    t_4 < t < t_5,
		\end{cases}
	\end{gathered}
	\label{cnot_signal}
\end{equation}
where
\begin{equation}
	\begin{aligned}
		& t_1 = T_1,	\\
		& t_2 = T_1 + \frac{T_\text{c}}{2},	\\
		& t_3 = T_1 + \frac{T_\text{c}}{2} + T_2,	\\
		& t_4 = T_1 + T_\text{c} + T_2,	\\
		& t_5 = 2T_1 + T_\text{c} + T_2.
	\end{aligned}
	\notag
\end{equation}

As in the case of a single-qubit and iSWAP gate, we compute the time dependence of the adiabatic energy levels [see Fig.~\ref{Fig:cnot_energy_dynamics}(a)], introduce the values of the transition probabilities $\mathcal{P}_i$ for each diabatic transition $N_i$, and define all phase gains $\zeta^{(ij)}_k$ \eqref{zeta_multi} between adiabatic levels $E_i$ and $E_j$ for the various periods of the adiabatic evolution.

The matrices for the \textit{diabatic} transitions are defined as
\begin{equation}
	\begin{gathered}
		N_k	=\begin{pmatrix}
			1 & 0 & 0 & 0 \\
			0 & 1 & 0 & 0 \\
			0 & 0 & R_k e^{i  \phi_\text{Sk}} &  - \alpha_k T_k	\\
			0 & 0 & \alpha_k T_k & R_k e^{-i  \phi_\text{Sk}}  \end{pmatrix},
	\end{gathered}	
\end{equation}
where $k=1,2$, $\alpha_1 = 1$, $\alpha_2 = -1$.
The matrix for the \textit{adiabatic} evolution is given by Eq.~\eqref{U_adiabatic_evolution}.
The evolution matrix for the whole period can be found as \eqref{total_evolution_matrix}.

\begin{figure}[t]
	\centering{%
		\includegraphics[width=1	\columnwidth]{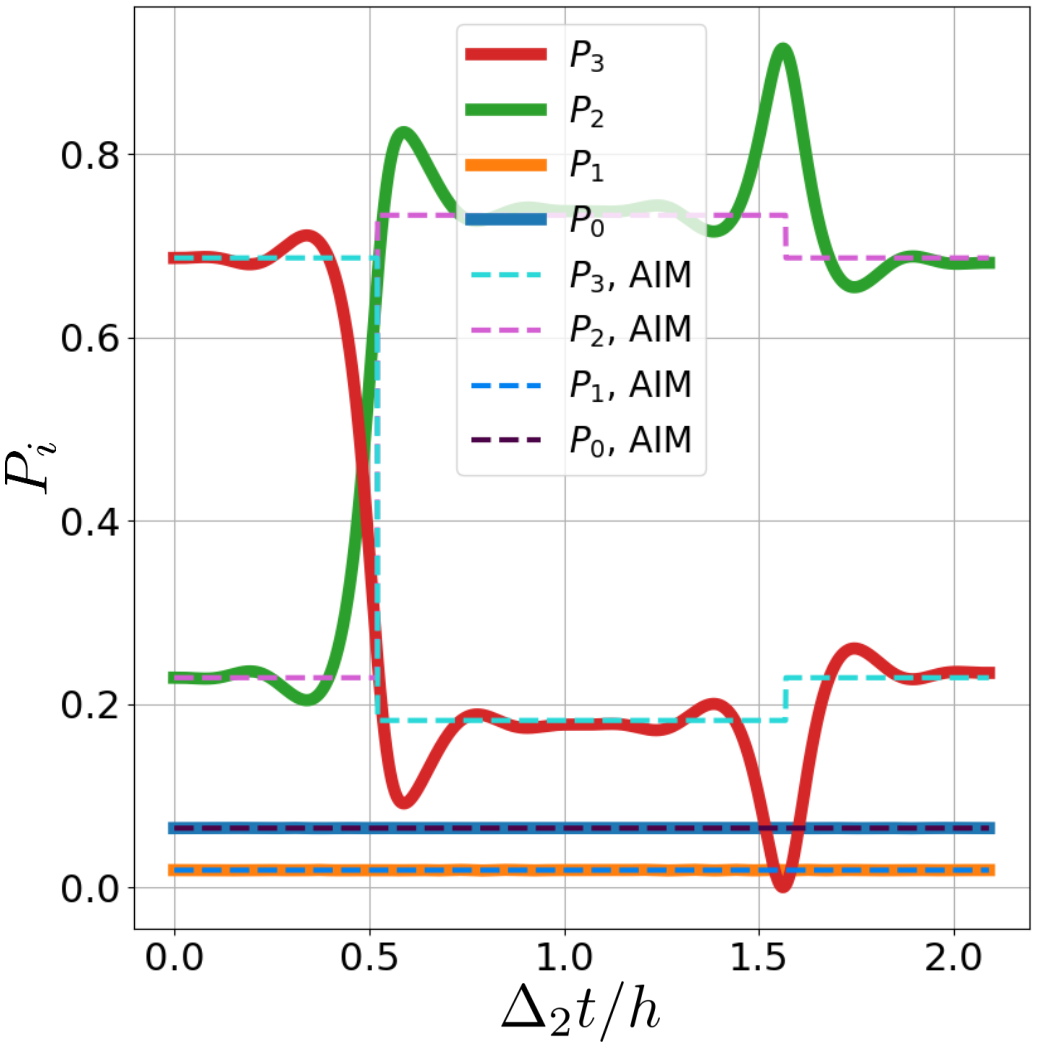}}
	\caption{  The dynamics of the CNOT gate implemented with two LZSM transitions. The occupation probabilities $P_i$ of each adiabatic level $E_i$ as function of time are obtained by two methods: numerical solutions of the Liouville-von Neumann equation and by the adiabatic-impulse model.
		The CNOT gate results in the swap of occupation probabilities between the $E_2$ and $E_3$ adiabatic energy levels.} 
	\label{Fig:cnot_dynamics}
\end{figure}
After simplifying by
\begin{equation}
	\begin{gathered}
		\zeta^{(01)} = \zeta^{(01)}_1 + \zeta^{(01)}_2 + \zeta^{(01)}_3,	\\
		\zeta^{(12)} = \zeta^{(12)}_1 + \zeta^{(12)}_2 + \zeta^{(12)}_3,
	\end{gathered}
\end{equation}
taking the common phase $e^{i \varphi}$ out of brackets and neglecting it (as the common phase of the wave function is irrelevant), we obtain the evolution matrix in the form
\begin{equation}
	\begin{gathered}
		\Xi = \begin{pmatrix} 
			1 & 0 & 0 & 0 \\
			0 & U_{11} & 0 & 0 \\
			0 & 0 & U_{22} & U_{23} \\
			0 & 0 & U_{32} & U_{33}    \end{pmatrix},
	\end{gathered}
	\label{cnot_evolution}
\end{equation}
which depends on the values $\mathcal{P}_1, \mathcal{P}_2, \zeta^{(01)}, \zeta^{(02)}, \zeta^{(23)}_i$.
Equating it to the matrix of a required two-qubit CNOT gate
\begin{equation}
	\begin{gathered}
		\text{CNOT}	=\begin{pmatrix} 
			1 & 0 & 0 & 0 \\
			0 & 1 & 0 & 0 \\
			0 & 0 & 0 & 1 \\
			0 & 0 & 1 & 0  \\  \end{pmatrix}
	\end{gathered}	\label{cnot}
\end{equation}
allows to determine the parameters of the external signal which implements this gate:
\begin{equation}
	\begin{aligned}
		&\mathcal{P}_1+\mathcal{P}_2  = 1,	\\
		&\zeta^{(01)}  =  \pi n_1, \\
		&\phi_\text{S1} + \phi_\text{S2} + 2 \zeta^{(23)}_2  = \pi +  2 \pi n_2,	\\
		&\phi_\text{S1} + 2 \zeta^{(12)} + 2 \zeta^{(23)}_1 + 2 \zeta^{(23)}_2 = 2 \pi n_3, \\
		&\phi_\text{S2} + 2 \zeta^{(12)} + 2 \zeta^{(23)}_2 + 2 \zeta^{(23)}_3 = 2 \pi n_4.
	\end{aligned}	
\end{equation}
For the considered signal \eqref{cnot_signal} with $\mathcal{P}_1 = \mathcal{P}_2 = \mathcal{P}$ and $\zeta^{(23)}_1 = \zeta^{(23)}_3$ the conditions simplify to
\begin{equation}
	\begin{aligned}
		&\mathcal{P}  = \frac{1}{2},	\\
		&\zeta^{(01)}  =  \pi n_1, \\
		&\phi_\text{S} +  \zeta^{(23)}_2 = \frac{\pi}{2} +   \pi n_2,	\\
		&\phi_\text{S} + 2 \zeta^{(12)} + 2 \zeta^{(23)}_1 + 2 \zeta^{(23)}_2 = 2 \pi n_3.
	\end{aligned}	
\end{equation}
In Fig.~\ref{Fig:cnot_dynamics} we illustrate the dynamics of the CNOT gate for a particular solution with $\Delta_2 T_\text{c}/h=2.0394$, $\Delta_2 T_1/h=0.0109$,  $\Delta_2 T_2/h=0.0288$, $A/ \Delta_2 = 4.6217$ for the Hamiltonian \eqref{Hamiltonian} with the parameters $\Delta_1/g=0.3$,  $\Delta_2/g=1$, $\varepsilon_1/g=16.6$, $J/g=10$, and compare the approximate solution obtained by the adiabatic-impulse model with the numerical solution of the Liouville-von Neumann equation. The same parameters were used for Fig.~\ref{Fig:cnot_energy_dynamics}.

The evolution matrix \eqref{cnot_evolution} can also implement the CZ, CPHASE, and other CNOT-like gates.
	
\nocite{apsrev41Control} 
\bibliography{LZSM2,1}

\end{document}